\documentclass[prb,preprint]{revtex4-1} 

\usepackage{amsmath}  
\usepackage{amsfonts} 
\usepackage{graphicx} 

\begin{document}


\title{Perihelion precession in binary systems: higher order corrections.}


\author{Jorge Poveda }
\email{jorge.okuden@gmail.com} 
\affiliation{Department of Physics, Universidad San Francisco de Quito, Diego de Robles y V\'ia Interoce\'anica} 


\author{Carlos Mar\'in}
\email{cmarin@usfq.edu.ec}
\affiliation{Department of Physics, Universidad San Francisco de Quito, Diego de Robles y V\'ia Interoce\'anica }


\date{\today}
\begin{abstract}{Higher order corrections (up to n-th order) are obtained for the perihelion precession in binary systems like OJ287 using the Schwarzschild metric and complex integration. The corrections are performed considering the third root of the motion equation and developing the expansion in  terms of $r_s/\left(a(1-e^2)\right)$.}The results are compared  with other expansions that appear in the literature giving corrections to second and third order. Finally, we simulate the shape of relativistic orbits for binary systems with different masses.

\end{abstract}


\maketitle 
\section{Introduction}
\label{intro}
Between 1908 and 1915, Albert Einstein made several unsuccessful attempts to obtain a theory of gravitation that was compatible with the Special Theory of Relativity (1905). In November 1915 he finally succeeded and called it "General Theory of Relativity" (GTR).
For the formulation of the GTR, Einstein relied on the principle of equivalence between inertial mass  and gravitational mass, which in turn implies the inability to distinguish between acceleration and gravity. Einstein realized that this equivalence could only be maintained if there was a connection between the gravitational force and the geometry of space.

The General Theory of Relativity is expresed in 14 equations \cite{Hobson,CMarin,Marin2}, the ten field equations:

\begin{eqnarray}
G_{\mu \nu}  \equiv R_{\mu \nu} - \frac{1}{2} R g_{\mu \nu} = k T_{\mu \nu} + \lambda g_{\mu \nu}
\label{eq:EcCamEinst}
\end{eqnarray}

and the geodesic equations (4 equations)
\begin{eqnarray}
\frac{d^{2}x^{\mu}}{ds^{2}} + \Gamma^{\mu}_{\rho \sigma}(\frac{dx^{\rho}}{ds})
(\frac{dx^{\sigma}}{ds}) = 0  \label{eq:Geodes}
\end{eqnarray}

In  equation (\ref{eq:EcCamEinst})  $G_{\mu \nu}$ is the Einstein' s Tensor, which describes the curvature of space-time, $R_{\mu \nu}$ is the Ricci tensor, and $R$ is the Ricci scalar (the trace of the Ricci tensor), $g_{\mu \nu}$ is the metric tensor that describes the deviation of the Pythagoras theorem in a curved space, $T_{\mu \nu}$ is the stress-energy tensor describing the content of matter and energy. $k = \frac{ 8 \pi G}{c^{4}}$, where $c = 299792458$ is the speed of light in vacuum and $G = 6.67384(80)\times10^{-11}$ is the gravitational constant. Finally, $\lambda$ is the cosmological constant introduced by Einstein in 1917 \cite{Weinberg1,Weinberg2,Hobson} that is a measure of the contribution to  the energy density of the universe due to vacuum fluctuations ($\vert \lambda \vert < 3 \times 10^{-52} m^{-2}$). In equation (\ref{eq:Geodes}), $x^{\mu}$ are the space-time coordinates of the particle.  We use Greek letters as $\mu , \nu , \alpha, $etc for 0,1,2,3. We have adopted the Einstein summation convention in which we sum over repeated indices.  $\Gamma_{\rho\sigma}^{\mu}$ are the Christoffel symbols of second kind:

\begin{equation}
\Gamma_{\rho\sigma}^{\mu}=\frac{1}{2}g^{\mu\alpha}\left\{ \partial_{\sigma}g_{\rho\alpha}+\partial_{\rho}g_{\sigma\alpha}-\partial_{\alpha}g_{\rho\sigma}\right\} 
\end{equation}

Finally  $s$ is the arc length satisfying the relation $ds^{2} = g_{\mu \nu}dx^{\mu}dx^{\nu}$. 
Einstein's equations (\ref{eq:EcCamEinst},\ref{eq:Geodes})  tells us that the curvature of a region of  space-time is determined by the distribution of mass-energy of the same\cite{Hobson,CMarin,Einstein2}.

\begin{figure}
\begin{centering}
\includegraphics{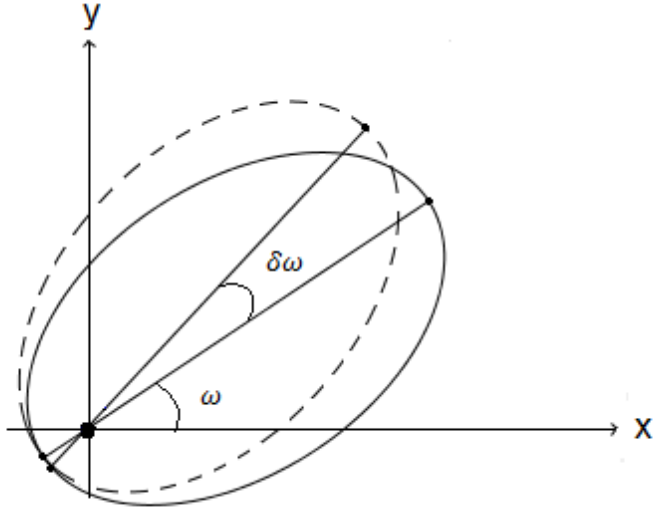}
\par\end{centering}
\caption{Perihelion precession. $\omega$ is the initial inclination of the orbit and $\delta\omega$ is the angle of precession.\label{fig:avance}}
\end{figure}

One of the most relevant predictions of General Relativity is the apsidal precession of elliptic orbits. Non circular orbits in GR are not perfect closed ellipses, but can be approximated to ellipses that precess, describing a patern like that of figure (\ref{fig:avance}) \cite{Carroll, Hartle, Will}. The perihelion precession can be computed using the Einstein's perihelion formula:

\begin{equation}
\delta\omega=\frac{6\pi GM}{a(1-e^{2})c^{2}}\label{eq:avanceclasic}
\end{equation}

The planet that presents the bigger precession is Mercury because of its proximity to the Sun. Now, it is difficult to compare the value of the precession with the experimental one because there are other factors that cause precession, as perturbations of other planets. Nevertheless, the relativistic contribution of the perihelion precession of Mercury is around 43 arc seconds per century and the remaining is due to different kind of perturbations.

Equation (\ref{eq:avanceclasic}) was deduced by Einstein using GR and with many approximations. It was successful to predict the values of the perihelion precession of the solar system planets, but it is not valid for strong gravitational fields, as the generated by a super massive black hole. In the literature one can find calculations of the perihelion precession using different methods and different theories as modified gravity \cite{Fokas, Tyler, Rosales}. Also, one can find corrections to second and third order of the perihelion precession using Schwarzschild coordinates \cite{Biesel, DEliseo, Scharf, Do_Nhat}. Higher order corrections (up to n-th order) are obtained in this paper using the Schwarzschild metric and complex integration. The corrections are performed in terms of the third root of the motion equation and not in terms of $r_s/a$ , where $r_{s}$  is the Schwarzschild radius.

\section{Schwarzschild metric}
\label{Schwarzschild}

In 1916 Karl Schwarzschild found the first exact solution to the Einstein field equations. For a spherical symmetric space-time with a mass $M$ in the center of the coordinate system, the invariant interval is \cite{Misner,Kenyon}:
\begin{eqnarray}
\left(ds\right)^{2} = \gamma\left(c dt\right)^{2} - \gamma^{-1}\left(dr\right)^{2} -r^{2}\left(d\Omega \right)^{2}  \label{eq:MetricSchwar}
\end{eqnarray}
where $\left( d \Omega \right)^{2} = \left(d \theta\right)^{2} +sin^{2}\theta \left(d \phi\right)^{2}$ , with coordinates  $x^{0}=ct$, $x^{1}=r$, $ x^{2}=\theta$ 
and  $x^{3}=\phi$. $\gamma = 1 - \frac{r_{s}}{r}$ where $ r_{s} = \frac{2GM}{c^{2}}$
is the Schwarzschild radius.

Then, the covariant metric tensor is:

\begin{equation}
g_{\mu\nu}=\left(\begin{array}{cccc}
\gamma & 0 & 0 & 0\\
0 & -\gamma^{-1} & 0 & 0\\
0 & 0 & -r^{2} & 0\\
0 & 0 & 0 & -r^{2}\sin^{2}\theta
\end{array}\right)
\end{equation}  \\

There are two singularities in this metric. The first when $\gamma=0$ or $r= r_{s}$ is a mathematical singularity that can be removed  by a convenient coordinate transformation like the one  introduced by Eddington in 1924 or Finkelstein in1958
\cite{Marin2,Kenyon}:

\begin{equation}
\hat{t}=t\pm\frac{r_{s}}{c}ln\left|\frac{r}{r_{s}}-1\right|  \label{eq:Eddington}
\end{equation}
With this coordinate transformation the invariant interval can be written as:

\begin{eqnarray}
\left(ds\right)^{2} = c^{2}\left(1 - \frac{r_{s}}{r}\right)\left(d \hat{t} \right)^{2} - 
\left(1 + \frac{r_{s}}{r}\right) \left(dr\right)^{2} \mp 2c \left(\frac{r_{s}}{r}\right) d\hat{t} dr
- r^{2} \left(d \Omega\right)^{2}
\end{eqnarray}
The first transformation in equation (\ref{eq:Eddington}) 
\begin{equation}
\hat{t}=t + \frac{r_{s}}{c}ln\left|\frac{r}{r_{s}}-1\right|  \label{eq:Eddington1}
\end{equation}
describes a black hole, while the second one:
\begin{equation}
\hat{t}=t - \frac{r_{s}}{c}ln\left|\frac{r}{r_{s}}-1\right|  \label{eq:Eddington2}
\end{equation}
represents what physicists call a "white hole" emitting material from a singularity in $r=0$ toward space-time.

The other singularity $r=0$ is physical, so it can not be removed. In this singularity, all known physical laws fail, and the curvature of space-time is infinite. If one particle reaches the event horizon ($r=r_s$), it will eventually falls to the singularity $r=0$, and it will never escape from the black hole neglecting quantum effects like Hawking radiation.

\section{Equation of motion}
\label{equation of motion}

Lets consider a gravitational source of mass $M$ and a massive particle that moves around the other. The motion of such particle is governed by the Schwarzschild metric. It is known that the orbital motion of celestial bodies is performed in a single plane because the orbital angular momentum must be constant. Then, we can analyze without problem the motion in the plane $\theta=\pi/2$ (equatorial plane):

\begin{equation}
(ds)^{2} = c^{2}(d\tau)^{2}=\gamma c^{2}(dt)^{2}-\gamma^{-1}(dr)^{2}-r^{2}(d\phi)^{2}\label{eq:Seg2}
\end{equation}

The geodesic equation can be written in an alternative form using the Lagrangian

\begin{eqnarray}
L\left(x^{\mu}, \frac{dx^{\mu}}{d\sigma}\right) = - g_{\alpha \beta}\left(x^{\mu}\right)
\frac{dx^{\alpha}}{d\sigma}\frac{dx^{\beta}}{d\sigma}
\end{eqnarray}
where $\sigma$ is a parameter of the trajectory of the particle, which is usually taken to be the proper time, $\tau$ for a massive particle. The resulting geodesic equation is:

\begin{eqnarray}
\frac{du_{\mu}}{d\tau} = \frac{1}{2} \left(\partial_{\mu} g_{\alpha \beta}\right) u^{\alpha} u^{\beta}   \label{eq:geodesic2}
\end{eqnarray}
where $u_{\mu} = \frac{dx_{\mu}}{d \tau}$.
For the coordinates $ct$ ($\mu = 0$) and  $\phi$ ($\mu =3$) the geodesic equation  (\ref{eq:geodesic2}) give us, respectively :

\begin{eqnarray}
\frac{d}{d \tau} \left[ \gamma c^{2} \frac{dt}{d \tau}\right] = 0
\end{eqnarray}
and
\begin{eqnarray}
\frac{d}{d \tau}\left[r^{2} \frac{d \phi}{d \tau} \right] = 0
\end{eqnarray}

This implies that there are two constants of motion:

\begin{equation}
E'=c^{2}\gamma\frac{dt}{d\tau}\label{eq:energia}
\end{equation}
and

\begin{equation}
J=r^{2}\frac{d\phi}{d\tau}\label{eq:momang}
\end{equation}

The first constant is the energy per unit mass, meanwhile the second is the angular momentum per unit mass.

With these constants of motion, we can rewrite the Schwarzschild metric to obtain:

\begin{eqnarray}
\left(\frac{dr}{d\tau}\right)^{2}=A+\frac{2GM}{r}-\frac{J^{2}}{r^{2}}\gamma \label{eq:mov}
\end{eqnarray}
and because 
\begin{equation}
\frac{dr}{d\tau} = \left(\frac{dr}{d\phi}\right)\left(\frac{d\phi}{d\tau}\right)
\end{equation}
we have additionally
\begin{equation}
\left(\frac{dr}{d\phi}\right)^{2}=\frac{A}{J^{2}}r^{4}-\gamma r^{2}+\frac{2GM}{J^{2}}r^{3}\label{eq:mov2}
\end{equation}
where $A = E'^{2}/c^{2} - c^{2}$.
The first equation  relates the radial distance $r$  with the proper time $\tau$, while the second equation relates the angle $\phi$ and the radial distance $r$.

\section{Constants and roots of the equation of motion}
\label{constants and roots}

Let's  consider both equations (\ref{eq:mov}) and (\ref{eq:mov2}).
If the path described by the particle is an ellipse there are two points where the radial velocity is zero. These points are the aphelion and the perihelion, and  satisfy: 

\[
A+\frac{2GM}{R_{a}}-\frac{J^{2}}{R_{a}^{2}}\gamma_{a}=0
\]

\[
A+\frac{2GM}{R_{p}}-\frac{J^{2}}{R_{p}^{2}}\gamma_{p}=0
\]
where $R_a$ is the distance to the aphelion and $R_p$ is the distance to the perihelion. Expanding $\gamma_{a}$ and $\gamma_{p}$:

\begin{equation}
A+\frac{2GM}{R_{a}}-\frac{J^{2}}{R_{a}^{2}}+\frac{J^{2}r_{s}}{R_{a}^{3}}=0\label{eq:aphe}
\end{equation}

\begin{equation}
A+\frac{2GM}{R_{p}}-\frac{J^{2}}{R_{p}^{2}}+\frac{J^{2}r_{s}}{R_{p}^{3}}=0\label{eq:perihe}
\end{equation}

Subtracting both equations:

\[
\frac{2GM}{R_{a}}-\frac{2GM}{R_{p}}=J^{2}\left(\frac{1}{R_{a}^{2}}-\frac{1}{R_{p}^{2}}-\frac{r_{s}}{R_{a}^{3}}+\frac{r_{s}}{R_{p}^{3}}\right)
\]

\[
2GM=J^{2}\left(\frac{R_{p}+R_{a}}{R_{a}R_{p}}-r_{s}\frac{R_{p}^{2}+R_{p}R_{a}+R_{a}^{2}}{R_{a}^{2}R_{p}^{2}}\right)
\]

Using the definitions $R_{a}=(1+e)a$ and $R_{p}=(1-e)a$:

\[
2GM=J^{2}\left(\frac{2a}{\left(1-e^{2}\right)a^{2}}-r_{s}\frac{\left(1-e\right)^{2}+1-e^{2}+\left(1+e\right)^{2}}{\left(1-e^{2}\right)^{2}a^{2}}\right)
\]

Where $a$ is the semi-major axis and $e$ is the eccentricity of the orbit. Simplifying the last equation:

\[
2GM\left(1-e^{2}\right)a^{2}=J^{2}\left(2a-r_{s}\frac{3+e^{2}}{1-e^{2}}\right)
\]

Finally, the total angular momentum would be:

\begin{equation}
J^{2}=\frac{GM\left(1-e^{2}\right)a}{1-\frac{r_{s}}{2a}\frac{3+e^{2}}{1-e^{2}}} \label{eq:angmom}
\end{equation}

Doing a Taylor's expansion in function of $\frac{r_{s}}{2a}\frac{3+e^{2}}{1-e^{2}}$, the angular momentum can be written as:

\begin{equation}
J^{2}=GM\left(1-e^{2}\right)a \sum_{n=0}^\infty \left(\frac{r_{s}}{2a}\frac{3+e^{2}}{1-e^{2}}\right)^n \label{eq:angmom1}
\end{equation}

In the Newtonian limit $r_s\ll a$, and then $\frac{r_{s}}{2a}\ll 1$. Furthermore, for classic orbits the eccentricity is usually small compared to 1, and then we can take $\frac{r_{s}}{2a}\frac{3+e^{2}}{1-e^{2}}\ll 1$. Taking only the first term:

\[
J^{2}\approx GM\left(1-e^{2}\right)a
\]

This is the classical expression of angular momentum. To have a better accuracy in the calculation of $J$, it can be taken the other terms depending on the value of $\frac{r_{s}}{2a}\frac{3+e^{2}}{1-e^{2}}$.

To obtain the energy, we can replace the expression of the angular momentum in equation (\ref{eq:aphe}):

\[
A+\frac{2GM}{R_{a}}-\frac{GM\left(1-e^{2}\right)a}{R_{a}^{2}\left(1-\frac{r_{s}}{2a}\frac{3+e^{2}}{1-e^{2}}\right)}+\frac{GM\left(1-e^{2}\right)ar_{s}}{R_{a}^{3}\left(1-\frac{r_{s}}{2a}\frac{3+e^{2}}{1-e^{2}}\right)}=0
\]
Using the expression of $R_{a}$:
\[
A=-\frac{2GM}{\left(1+e\right)a}+\frac{GM\left(1-e\right)}{\left(1+e\right)a\left(1-\frac{r_{s}}{2a}\frac{3+e^{2}}{1-e^{2}}\right)}-\frac{GM\left(1-e\right)r_{s}}{\left(1+e\right)^{2}a^{2}\left(1-\frac{r_{s}}{2a}\frac{3+e^{2}}{1-e^{2}}\right)}
\]
and simplifying, we finally get:
\begin{equation}
A=\frac{GM\left[2r_{s}-\left(1-e^{2}\right)a\right]}{\left(1-e^{2}\right)a^{2}\left(1-\frac{r_{s}}{2a}\frac{3+e^{2}}{1-e^{2}}\right)} \label{eq:energ}
\end{equation}

With these constants of motion we can continue with the calculation of the perihelion precession. For this, lets rewrite equation (\ref{eq:mov2}) as:

\[
\left(\frac{dr}{d\phi}\right)^{2}J^{2}=Ar^{4}-\gamma J^{2}r^{2}+2GMr^{3}
\]

Using the expression of $\gamma$ and changing $2GM$ by $r_{s}c^{2}$:

\begin{equation}
\left(\frac{dr}{d\phi}\right)^{2}=r\left(\frac{A}{J^{2}}r^{3}+\frac{r_{s}c^{2}}{J^{2}}r^{2}-r+r_{s}\right)
\end{equation}

For an ellipse, the equation of motion have three real and positive roots. Two of the roots are $R_a$ and $R_p$ and the other we will call $R_{0}$. The third root can be calculated multiplying the factors of the equation and comparing them. For this it is important to recall that $A$ is negative for elliptic orbits, so it can be written as $A=-\left|A\right|$. Then:

\begin{equation}
-\frac{\left|A\right|}{J^{2}}r^{3}+\frac{r_{s}c^{2}}{J^{2}}r^{2}-r+r_{s}=\frac{\left|A\right|}{J^{2}}\left(R_{a}-r\right)\left(r-R_{p}\right)\left(r-R_{o}\right)
\end{equation}

In the first factor we wrote $R_{a}-r$, because $R_{a}$ is the maximum value that $r$ can take. Multiplying and simplifying:

\[
-\frac{\left|A\right|}{J^{2}}r^{3}+\frac{r_{s}c^{2}}{J^{2}}r^{2}-r+r_{s}=\frac{\left|A\right|}{J^{2}} (-r^{3}+r^{2}R_{o}+r^{2}R_{p}-rR_{p}R_{o}\qquad\qquad\qquad\qquad\]

\[\qquad\qquad\qquad\qquad\qquad\qquad\qquad\qquad+R_{a}r^{2}-R_{a}R_{p}r-R_{a}R_{o}r+R_{a}R_{p}R_{o})
\]

\[
\frac{r_{s}c^{2}}{J^{2}}r^{2}-r+r_{s}=\frac{\left|A\right|}{J^{2}}\left(R_{o}+R_{a}+R_{p}\right)r^{2}-\frac{\left|A\right|}{J^{2}}\left(R_{p}R_{o}+R_{a}R_{p}+R_{a}R_{o}\right)r+\frac{\left|A\right|}{J^{2}}R_{a}R_{p}R_{o}
\]

In the last equation, the coefficients of the powers of $r$ must be the same by linear independence, so we have three new equations:

\[
\frac{\left|A\right|}{J^{2}}\left(R_{o}+R_{a}+R_{p}\right)=\frac{r_{s}c^{2}}{J^{2}}
\]

\[
\frac{\left|A\right|}{J^{2}}\left(R_{p}R_{o}+R_{a}R_{p}+R_{a}R_{o}\right)=1
\]

\[
\frac{\left|A\right|}{J^{2}}R_{a}R_{p}R_{o}=r_{s}
\]

Replacing the values of $R_{a}$ and $R_{p}$:

\[
R_{o}+2a=\frac{r_{s}c^{2}}{\left|A\right|}
\]

\[
2aR_{o}+\left(1-e^{2}\right)a^{2}=\frac{J^{2}}{\left|A\right|}
\]

\[
\left(1-e^2\right)a^{2}R_{o}=\frac{r_{s}J^{2}}{\left|A\right|}
\]

To obtain $R_o$ we use the last two equations:

\[
\left(1-e^2\right)a^{2}R_{o}=2ar_{s}R_{o}+\left(1-e^{2}\right)r_{s}a^{2}
\]
simplifying we get:
\begin{equation}
R_{o}=\frac{\left(1-e^{2}\right)r_{s}a}{\left(1-e^2\right)a-2r_{s}} \label{eq:raiz}
\end{equation}
For classical systems $a\gg r_s$. So using this fact, the third root can be reduced to:
\[
R_{o}=\left(\frac{\left(1-e^{2}\right)r_{s}}{\left(1-e^2\right)-2\frac{r_{s}}{a}}\right)\approx r_{s}
\]
Then $R_{o}$ is of the order of $r_{s}$ in this limit, and consequently $R_{o}\ll R_{p}<R_{a}$. Nevertheless, $R_{o}$ is always smaller than $R_{p}$ and $R_{a}$.

With the third root we can rewrite equation  (\ref{eq:mov}) as:

\begin{equation}
\left(\frac{dr}{d\phi}\right)^{2}=\frac{\left|A\right|}{J^{2}}\left(R_{a}-r\right)\left(r-R_{p}\right)\left(r-R_{o}\right)r
\end{equation}

To find the angle travelled in a period, we can integrate the last equation from the perihelion position to the aphelion position and multiply it by two. Classically, the value of such angle must be $2\pi$, but in this case there is a little deviation. This deviation is the perihelion precession.

\begin{equation}
\Delta\phi=\frac{2J}{\left|A\right|^{1/2}}\intop_{R_{p}}^{R_{a}}\frac{dr}{\left[\left(R_{a}-r\right)\left(r-R_{p}\right)\left(r-R_{o}\right)r\right]^{1/2}} 
\end{equation}

or

\begin{eqnarray}
\Delta\phi=\frac{2J}{\left|A\right|^{1/2}}\intop_{R_{p}}^{R_{a}}\frac{r^{-1/2}\left(1-\frac{R_{o}}{r}\right)^{-1/2}dr}{\left[\left(R_{a}-r\right)\left(r-R_{p}\right)r\right]^{1/2}} \label{eq:angle}
\end{eqnarray}

It was shown that $R_{o}$ is smaller than $R_{p}$, so it must be smaller than the radial distance $r$ for all $t$. This allows us to expand equation (\ref{eq:angle}) in a power series around $R_o$:

Then, the angle would be:

\begin{equation}
\Delta\phi=\frac{2J}{\left|A\right|^{1/2}}\sum_{n=1}^{\infty}\left(\begin{array}{c}
-\frac{1}{2}\\
n-1
\end{array}\right)(-1)^{n-1}R_{o}^{n-1}I_{n} \label{eq:apsidal}
\end{equation}
with

\begin{equation}
I_{n}=\intop_{R_{p}}^{R_{a}}\frac{dr}{r^{n}\left[\left(R_{a}-r\right)\left(r-R_{p}\right)\right]^{1/2}}
\end{equation}

Using the values of $R_{a}$ and $R_{p}$:

\begin{equation}
I_{n}=\intop_{R_{p}}^{R_{a}}\frac{dr}{r^{n}\left[e^{2}a^{2}-\left(r-a\right)^{2}\right]^{1/2}}
\end{equation}

To solve this integral we can use a change of variable: $r-a=eacos\theta$, such that the integral takes the form:

\begin{equation}
I_{n}=\frac{1}{a^{n}}\intop_{0}^{\pi}\frac{d\theta}{\left(1+ecos\theta\right)^{n}}
\end{equation}

\begin{figure}
\begin{centering}
\includegraphics[scale=0.7]{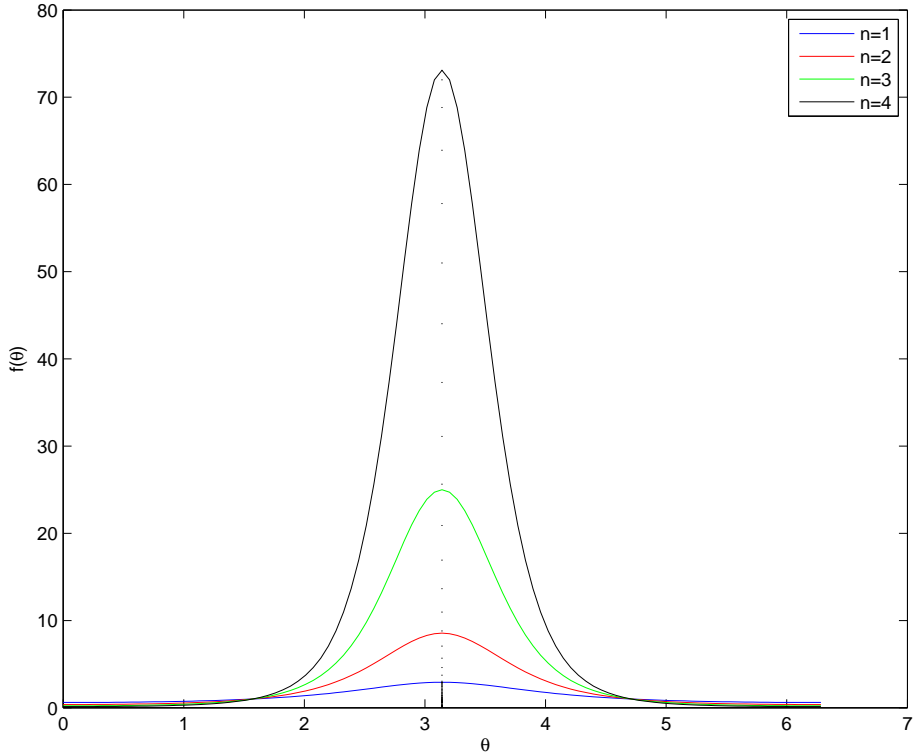}
\par\end{centering}

\caption{Plot of $f(\theta)=\frac{1}{\left(1+ecos\theta\right)^{n}}$ for different values of $n$ around $\theta=\pi$.\label{fig:avance-1}}
\end{figure}

As it can be seen in figure (\ref{fig:avance-1}), the function $(1+ecos\theta)^{-n}$ is symmetric around $\theta=\pi$. This allows us to write:

\begin{equation}
I_{n}=\frac{1}{2a^{n}}\intop_{0}^{2\pi}\frac{d\theta}{\left(1+ecos\theta\right)^{n}}
\end{equation}

Lets use the substitution $cos\theta=\frac{1}{2}\left(e^{i\theta}+e^{-i\theta}\right)$, and define $z=e^{i\theta}$ with $\left|z\right|=1$. Then:

\begin{equation}
\cos\theta=\frac{1}{2}\left(z+\frac{1}{z}\right)=\frac{1}{2}\left(\frac{z^{2}+1}{z}\right)
\end{equation}
and

\begin{equation}
d\theta=\frac{dz}{iz}
\end{equation}

So, we have to calculate the integral:

\begin{equation}
I_{n}=\frac{2^{n-1}}{ia^{n}e^{n}}\ointop\frac{z^{n-1}dz}{\left(z^{2}+\frac{2}{e}z+1\right)^{n}}
\end{equation}

Calculating the factors of the denominator, we can find the poles of the function:

\begin{equation}
z^{2}+\frac{2}{e}z+1=e\left(z+\frac{1+\sqrt{1-e^{2}}}{e}\right)\left(z+\frac{1-\sqrt{1-e^{2}}}{e}\right)
\end{equation}

Then the poles are:

\begin{equation}
z_{1}=-\frac{1+\sqrt{1-e^{2}}}{e}
\end{equation}
and

\begin{equation}
z_{2}=-\frac{1-\sqrt{1-e^{2}}}{e}
\end{equation}

Before we continue with the calculations, lets list some properties of the poles: $z_{1}+z_{2}=-\frac{2}{e}$, $z_{1}-z_{2}=-\frac{2\sqrt{1-e^{2}}}{e}$ and $z_{1}z_{2}=1$.

With the poles we can rewrite the integral as:

\begin{equation}
I_{n}=\frac{2^{n-1}}{ia^{n}e^{n}}\ointop\frac{z^{n-1}dz}{\left(z-z_{1}\right)^{n}\left(z-z_{2}\right)^{n}}=\frac{2^{n-1}}{ia^{n}e^n}\ointop f(z)dz\label{eq:I-1}
\end{equation}
with

\begin{equation}
f(z)=\frac{z^{n-1}}{\left(z-z_{1}\right)^{n}\left(z-z_{2}\right)^{n}}
\end{equation}

The path of integration is the unitary circumference, because $\left| z \right|=1$. Now, as $0<e<1$, the first pole is $\left|z_{1}\right|\geq 1$. So, $z_{1}$ is out of the integration zone and it is not important. The second pole is $\left|z_{2}\right|\leq 1$. Then, using the residue theorem:

\begin{equation}
\oint f(z)dz=2\pi i Res_{z=z_{2}}f(z)
\end{equation}

The residue can be calculated by:

\[
Res_{z=z_{2}}f(z)=\lim_{z\rightarrow z_{2}}\frac{1}{\left(n-1\right)!}\frac{d^{n-1}}{dz^{n-1}}\left(\left(z-z_{2}\right)^{n}f(z)\right)
\]

Calculating this:

\[
Res_{z=z_{2}}f(z)=\left.\frac{1}{\left(n-1\right)!}\frac{d^{n-1}}{dz^{n-1}}\left[\frac{z^{n-1}}{\left(z-z_{1}\right)^{n}}\right]\right|_{z=z_{2}}
\]

And then the value of the integral would be:

\begin{equation}
I_{n}=\frac{2^{n}\pi}{a^{n}e^{n}\left(n-1\right)!}\left.\frac{d^{n-1}}{dz^{n-1}}\left[\frac{z^{n-1}}{\left(z-z_{1}\right)^{n}}\right]\right|_{z=z_{2}}\label{eq:integ}
\end{equation}

Now lets compute some of the first terms. For $n=1$:

\begin{equation}
I_{1}=\frac{2\pi}{ae}\left[\frac{1}{\left(z_2-z_{1}\right)}\right]
\end{equation}

Replacing the values of $z_{1}$ y $z_{2}$:

\begin{equation}
I_{1}=\frac{\pi}{a\left(1-e^{2}\right)^{1/2}}
\end{equation}

For $n=2$:

\[
I_{2}=\frac{4\pi}{a^{2}e^{2}}\left.\frac{d}{dz}\left[\frac{z}{\left(z-z_{1}\right)^{2}}\right]\right|_{z=z_{2}}=-\frac{4\pi}{a^{2}e^{2}\left(z_{2}-z_{1}\right)^{3}}\left[z_{1}+z_{2}\right]
\]

Replacing the values of $z_{1}$ and $z_{2}$:

\begin{equation}
I_{2}=\frac{\pi}{a^{2}\left(1-e^{2}\right)^{3/2}}
\end{equation}

For $n=3$:

\[
I_{3}=\frac{2^{2}\pi}{a^{3}e^{3}}\left.\frac{d^{2}}{dz^{2}}\left[\frac{z^{2}}{\left(z-z_{1}\right)^{3}}\right]\right|_{z=z_{2}}=\frac{2^{2}\pi}{a^{3}e^{3}}\frac{2\left(z_{1}^{2}+4z_{1}z_{2}+z_{2}^{2}\right)}{\left(z_{2}-z_{1}\right)^5} 
\]

Replacing the values of $z_{1}$ y $z_{2}$ we get:

\begin{equation}
I_{3}=\frac{\pi}{a^{3}\left(1-e^{2}\right)^{5/2}}\left(1+\frac{e^{2}}{2}\right)
\end{equation}

For $n=3$:

\[
I_{4}=\frac{2^{4}\pi}{6a^{4}e^{4}}\left.\frac{d^{3}}{dz^{3}}\left[\frac{z^{3}}{\left(z-z_{1}\right)^{4}}\right]\right|_{z=z_{2}}=\frac{2^{4}\pi}{6a^{4}e^{4}}\left(\frac{-6}{\left(z_{2}-z_{1}\right)^{7}}\left[z_{1}^{3}+9z_{1}^{2}z_{2}+9z_{1}z_{2}^{2}+z_{1}^{3}\right]\right)
\]

that reduces to

\begin{equation}
I_{4}=\frac{\pi}{a^{4}\left(1-e^{2}\right)^{7/2}}\left(1+\frac{3}{2}e^{2}\right)
\end{equation}

In general, using the Leibniz's formula, we have:

\begin{eqnarray}
\left.\frac{d^{n-1}}{dz^{n-1}}\left[\frac{z^{n-1}}{\left(z-z_{1}\right)^{n}}\right]\right|_{z=z_{2}}=
\sum_{k=0}^{n-1}\left(-1\right)^{k}\left(\begin{array}{c}
n-1\\
k
\end{array}\right) \frac{\left(n+k-1\right)!}{k!} z_{2}^{k}\left(z_{2}-z_{1}\right)^{-n-k}
\nonumber
\end{eqnarray}
that can also be written as:

\begin{eqnarray}
\left.\frac{d^{n-1}}{dz^{n-1}}\left[\frac{z^{n-1}}{\left(z-z_{1}\right)^{n}}\right]\right|_{z=z_{2}}=\frac{\left(n-1\right)!(-1)^{n+1}}{\left(z_{2}-z_{1}\right)^{2n-1}}\sum_{k=0}^{n-1}\left(\begin{array}{c}
n-1\\
k
\end{array}\right)^{2}z_{1}^{n-1-k}z_{2}^{k}
\end{eqnarray}

Where $\left(\begin{array}{c}
n-1\\
k
\end{array}\right)= \frac{\left(n-1\right)!}{k!\left(n-1-k\right)!}$ represents the coefficients of the binomial expansion. Replacing this in equation (\ref{eq:integ}):

\[
I_{n}=\frac{2^{n}\pi}{a^{n}e^{n}}\frac{(-1)^{n+1}}{\left(z_{2}-z_{1}\right)^{2n-1}}\sum_{k=0}^{n-1}\left(\begin{array}{c}
n-1\\
k
\end{array}\right)^{2}z_{1}^{n-1-k}z_{2}^{k}
\]

Introducing  the values of $z_{1}$ and $z_{2}$  in the part before the summation sign:

\begin{equation}
I_{n}=\frac{\pi(-1)^{n+1}}{a^{n}2^{n-1}\left(1-e^{2}\right)^{n-1/2}}\sum_{k=0}^{n-1}\left(\begin{array}{c}
n-1\\
k
\end{array}\right)^{2}z_{1}^{n-1-k}z_{2}^{k}e^{n-1}
\end{equation}

At this point we can define the functions $Q_{n-1}\left(z_1,z_2\right)$ as:

\begin{equation}
Q_{n}\left(z_1, z_2\right)=\sum_{k=0}^{n}\left(\begin{array}{c}
n\\
k
\end{array}\right)^{2}z_1^{n-k}{z_2}^{k}e^n
\end{equation}

\begin{table}[h]
\caption{Values of the functions $Q_n$ \label{tab:AgujNegTabla-1}}

\smallskip{}

\centering{}%
\begin{tabular}{ll}
\hline
\hline 
\noalign{\vskip\doublerulesep}
Function & Expression\tabularnewline[\doublerulesep]
\hline
\noalign{\vskip\doublerulesep}
$Q_{0}$ & $1$\tabularnewline[\doublerulesep]
\noalign{\vskip\doublerulesep}
$Q_{1}$ & $-2$\tabularnewline[\doublerulesep]
\noalign{\vskip\doublerulesep}
$Q_{2}$ & $\left(4+2e^{2}\right)$\tabularnewline[\doublerulesep]
\noalign{\vskip\doublerulesep}
$Q_{3}$ & $-\left(8+12e^{2}\right)$\tabularnewline[\doublerulesep]
\noalign{\vskip\doublerulesep}
$Q_{4}$ & $\left(16+48e^{2}+6e^{4}\right)$\tabularnewline[\doublerulesep]
\noalign{\vskip\doublerulesep}
$Q_{5}$ & $-\left(32+160e^{2}+60e^{4}\right)$\tabularnewline[\doublerulesep]
\hline
\hline
\end{tabular}
\end{table}

In table (\ref{tab:AgujNegTabla-1}) it is shown the first five functions $Q_n$. Replacing the value of the integral in equation (\ref{eq:apsidal}):

\begin{equation}
\Delta\phi=\frac{2\pi J}{\left|A\right|^{1/2}}\sum_{n=1}^{\infty}\left(\begin{array}{c}
-\frac{1}{2}\\
n-1
\end{array}\right)\frac{R_{o}^{n-1}e^{n-1}}{a^{n}2^{n-1}\left(1-e^{2}\right)^{n-1/2}}Q_{n-1}\left(z_{1},z_2\right)\label{eq: ang}
\end{equation}

Using the expressions of $J^{2}$ and $A$ (equations (\ref{eq:angmom}) and (\ref{eq:energ})) it can be shown that:

\begin{equation}
\frac{J^{2}}{\left|A\right|}=\frac{\left(1-e^{2}\right)^{2}a^{2}}{\left[\left(1-e^{2}\right)-2\frac{r_{s}}{a}\right]}
\end{equation}

Replacing this and changing the sum index to begin the sum at $n=0$:

\begin{equation}
\Delta\phi=\frac{2\pi(1-e^{2})^{1/2}}{\left[\left(1-e^{2}\right)-2\frac{r_{s}}{a}\right]^{1/2}}\sum_{n=0}^{\infty}\left(\begin{array}{c}
-\frac{1}{2}\\
n
\end{array}\right)\frac{R_{o}^{n}e^{n}}{2^{n}a^{n}\left(1-e^{2}\right)^{n}}Q_{n}\left(z_{1},z_2\right)\label{eq: ang}
\end{equation}

Now, we can replace the expression of $R_o$:

\begin{equation}
\Delta\phi=\frac{2\pi(1-e^{2})^{1/2}}{\left[\left(1-e^{2}\right)-2\frac{r_{s}}{a}\right]^{1/2}}\sum_{n=0}^{\infty}\left(\begin{array}{c}
-\frac{1}{2}\\
n
\end{array}\right)\frac{e^{n}Q_{n}\left(z_{1},z_2\right)}{2^{n}\left[\left(1-e^{2}\right)-2\frac{r_{s}}{a}\right]^{n}}\left(\frac{r_s}{a}\right)^n\label{eq: ang}
\end{equation}

We can define the quantity $\epsilon=\frac{r_s}{\left(1-e^{2}\right)a}$, and then

\begin{equation}
\Delta\phi=\frac{2\pi}{\left[1-\frac{2r_{s}}{a\left(1-e^{2}\right)}\right]^{1/2}}\sum_{n=0}^{\infty}\left(\begin{array}{c}
-\frac{1}{2}\\
n
\end{array}\right)\frac{Q_{n}\left(z_{1},z_2\right)}{2^{n}\left[1-\frac{2r_{s}}{a\left(1-e^{2}\right)}\right]^{n}}\left[\frac{r_s}{\left(1-e^{2}\right)a}\right]^n \label{eq:deltaphi}
\end{equation}

\begin{equation}
\Delta\phi=\frac{2\pi}{\left(1-2\epsilon\right)^{1/2}}\sum_{n=0}^{\infty}\left(\begin{array}{c}
-\frac{1}{2}\\
n
\end{array}\right)\frac{Q_{n}\left(z_{1},z_2\right)}{2^{n}\left(1-2\epsilon\right)^{n}}\epsilon^n\label{eq:ang2}
\end{equation}

Equation (\ref{eq:ang2}) is a general form to compute the value of the perihelion precession at any order. To do that, first we must find the values of $Q_n$, and then we can expand the series.

\section{Expansion in terms of $\epsilon=\frac{r_s}{a(1-e^2)}$}
\label{expansion}

Equations (\ref{eq:deltaphi},\ref{eq:ang2}) are not an expansion in powers of $r_s/a$ because of the denominators. They are expansions in terms of $\epsilon=\frac{r_s}{a(1-e^2)}$. An expansion in terms of $r_s/a$ will converge slower, but if we can recover the first terms of such expansion, we will prove that equation (\ref{eq:ang2}) is correct.

So lets compute the first three terms of (\ref{eq:ang2}) to recover the expansion until second order on $\epsilon$.

\begin{eqnarray}
\Delta\phi^{(2)}
& = & \frac{2\pi}{\left(1-2\epsilon\right)^{1/2}}\left[Q_{0}\left(z_{1},z_2\right)- \frac{1}{2}\frac{Q_{1}\left(z_1,z_2\right)}{2\left(1-2\epsilon\right)}\epsilon + \frac{3}{8}\frac{Q_{2}\left(z_{1},z_2\right)}{2^2\left(1-2\epsilon\right)^2}\epsilon^2\right]\nonumber \\
& = & \frac{2\pi}{\left(1-2\epsilon\right)^{1/2}}\left[1+\frac{\epsilon}{2\left(1-2\epsilon\right)}
+ \frac{3\left(2+e^2\right)\epsilon^2}{16\left(1-2\epsilon\right)^2} \right]
\end{eqnarray}

Approximating until second order in $\epsilon$:

\begin{equation}
\Delta\phi^{(2)}\approx 2\pi\left[1+\frac{3}{2} \epsilon+\frac{27}{8}\epsilon^2+\frac{3}{16}e^{2}\epsilon^{2} + ..... \right] \label{eq:segundo}
\end{equation}

Finally:

\begin{equation}
\Delta\phi^{(2)}\approx 2\pi+3\pi\epsilon + \frac{3\left(18+e^2\right)\pi}{8}\epsilon^2
\end{equation}

As the perihelion precession is $\chi=\Delta\phi-2\pi$, if we replace  the value of $\epsilon$ to second order we can write:

\begin{equation}
\chi^{(2)}\approx \frac{6\pi GM}{a\left(1-e^2\right)c^2} + \frac{3\left(18+e^2\right)\pi G^2M^2}{2\left(1-e^2\right)^2a^2c^4}   \label{eq:chi2}
\end{equation}

The first term agrees with the one calculated by Einstein. Also, the second term agrees with the calculated by Scharf\cite{Scharf}, but not with those calculated by Nhat\cite{Do_Nhat} and D'Eliseo\cite{DEliseo}. 

Using the same procedure, any term of the expansion can be calculated  using equation (\ref{eq:ang2}) . For example to third order in $\epsilon$ we have:

\begin{eqnarray}
 \chi^{\left(3\right)} = \Delta \phi^{\left(3\right)}-2\pi \approx 3\pi\epsilon + \frac{3\left(18+e^2\right)\pi}{8}\epsilon^2+
\frac{45\left(6+e^{2}\right)\pi}{16}\epsilon^{3}
 \label{eq:deltaphi3}
 \end{eqnarray}
 and then
 \begin{eqnarray}
 \chi^{\left(3\right)} \approx \frac{6\pi GM}{a\left(1-e^2\right)c^2} + \frac{3\left(18+e^2\right)\pi G^2M^2}{2\left(1-e^2\right)^2a^2c^4} \nonumber \\ +\frac{45\left(6+e^{2}\right)\pi G^{2}M^{2}}{2\left(1-e^{2}\right)^{3}a^{3}c^{6}}
 \label{eq:chi3}
 \end{eqnarray}
 
The first five terms of the expansion of  $\chi$ in  $\epsilon$ powers are shown in table (\ref{tab:expansion}).

\begin{table}[h]
\caption{Expressions of the first five terms of the perihelion precession. \label{tab:expansion}}

\smallskip{}

\centering{}%
\begin{tabular}{ll}
\hline
\hline 
\noalign{\vskip\doublerulesep}
Term & Expression\tabularnewline[\doublerulesep]
\hline
\noalign{\vskip\doublerulesep}
$1$ & $3\pi\epsilon$\tabularnewline[\doublerulesep]
\noalign{\vskip\doublerulesep}
$2$ & $\frac{3\left(18+e^2\right)\pi}{8}\epsilon^2$\tabularnewline[\doublerulesep]
\noalign{\vskip\doublerulesep}
$3$ & $\frac{45\left(6+e^2\right)\pi}{16}\epsilon^3$\tabularnewline[\doublerulesep]
\noalign{\vskip\doublerulesep}
$4$ & $\frac{105\left(216+72e^2+e^4\right)\pi}{512}\epsilon^4$\tabularnewline[\doublerulesep]
\noalign{\vskip\doublerulesep}
$5$ & $\frac{189\left(648+370e^2+5e^4\right)\pi}{1024}\epsilon^5$\tabularnewline[\doublerulesep]
\hline
\hline
\end{tabular}
\end{table}
In general $\chi^{\left(n\right)} = \Delta \phi^{\left(n\right)}-2\pi$.

\section{Applications}
\label{applications}

To compare equations (\ref{eq:avanceclasic}) and (\ref{eq:chi2}), the perihelion precession for the interior planets of the solar system was calculated in arc seconds per century. The values are shown in table (\ref{tab:AgujNegTabla}).

\begin{table}[h]
\caption{Perihelion precession for the interior planets of the solar system. \label{tab:AgujNegTabla}}

\smallskip{}

\centering{}%
\begin{tabular}{lllll}
\hline 
\hline
Planet & $e$ & $a$ ($UA$) & $\delta\omega$ $('')$ & $\chi^{(2)}$ $('')$\tabularnewline
\hline  
Mercury & 0.20563069  & 0.387098 & 42.9307597 & 42.9307643\tabularnewline
Venus & 0.00677323 & 0.723327 & 8.59734793 & 8.59734826\tabularnewline
Earth & 0.01671123 & 1.000003 & 3.83432636 & 3.83432651\tabularnewline
Mars & 0.093315 & 1.523679 & 1.34837228 & 1.34837232\tabularnewline
\hline
\hline 
\end{tabular}
\end{table}

It can be seen that the correction is not relevant. It is clear then that for weak gravitational fields, the classical formula is sufficiently precise.

OJ 287 is a binary system that produces periodic outbursts. These outbursts have been detected for approximately 100 years. The first observation it was done, was through a photographic plate in 1891. This system is located 3.500 million light years from Earth, and is theorized that it is a binary system of black holes. In table (\ref{tab:Oj287}) are especified the data of the orbit of OJ287 \cite{PauliPihajoki,Valtonen}. It is important to recall that the measurements have been questioned because of the limited number of orbital companions of the system.

\begin{table}[h]
\caption{Data of the binary system OJ287. \label{tab:Oj287}}

\smallskip{}

\centering{}%
\begin{tabular}{ll}
\hline
\hline 
Parameter & Value\tabularnewline
\hline 
$M$ & $\left(1.83\pm0.01\right)\times10^{10} M_{\Theta}$\tabularnewline
$m$ & $\left(1.50\pm0.1\right)\times10^{8} M_{\Theta}$\tabularnewline
$e$ & $0.70\pm0.001$\tabularnewline
$a$ & $\left(11500\pm8\right)AU$\tabularnewline
$s$ & $0.313\pm0.08$\tabularnewline
$P$ & $12$  years \tabularnewline
$\chi_{exp}$ & $\left(39.1\pm0.1\right){^\circ}$\tabularnewline
\hline
\hline 
\end{tabular}
\end{table}
The experimental perihelion precession in a cycle is approximately $39^{\circ}$. For the calculation of this parameter to different orders, we used equation (\ref{eq:ang2}) until a given order $m$:

\begin{equation}
\Delta\phi^{m}=\frac{2\pi}{\left(1-2\epsilon\right)^{1/2}}\sum_{n=0}^{m}\left(\begin{array}{c}
-\frac{1}{2}\\
n
\end{array}\right)\frac{Q_{n}\left(z_{1},z_2\right)}{2^{n}\left(1-2\epsilon\right)^{n}}\epsilon^n
\end{equation}

In table (\ref{tab:AgujNegTabla-2}) are shown the values of the perihelion precession using equations (\ref{eq:avanceclasic}) and (\ref{eq:ang2}) for different orders. It can be seen that between the first and the second order there is a difference of approximately $5^{\circ}$. For higher orders, the difference is less than $1^{\circ}$. For $m =4$, the value of the perihelion precession begins to stabilize around $38.87^{\circ}$.

Equation (\ref{eq:ang2}) gives an important correction to the perihelion precession. The measured value is slightly higher. The calculation can be improved if it is considered the spin of the central black hole, and the gravitational radiation of the system.

\begin{table}[h]
\caption{Calculation of the perihelion precession for OJ287 binary system. \label{tab:AgujNegTabla-2}}

\smallskip{}

\centering{}%
\begin{tabular}{llllll}
\hline
\hline
Precession & Value \tabularnewline 
\hline

$\chi^{(1)}$ & $33.223^{\circ}$ \tabularnewline
$\chi^{(2)}$ & $37.948^{\circ}$ \tabularnewline
$\chi^{(3)}$ & $38.713^{\circ}$ \tabularnewline
$\chi^{(4)}$ & $38.846^{\circ}$ \tabularnewline
$\chi^{(5)}$ & $38.876^{\circ}$ \tabularnewline

\hline
\hline 
\end{tabular}
\end{table}

\section{Discussion about the different expansions}
\label{discussion}

In the works of D'Eliseo \cite{DEliseo} and Do Nhat \cite{Do_Nhat}, the first three terms of the expansion calculated are: 

\begin{equation}
\Delta\omega'^{\left(3\right)}=2\pi\epsilon'+\frac{5\pi\left(6+e'^2\right)}{6}\epsilon'^2+\frac{5\pi\left(54-6e'+15e'^2-2e'^3\right)}{18}\epsilon'^3 \label{eq:DoNhat}
\end{equation}
where $\epsilon'=\frac{3}{2}\frac{r_s}{(1-e'^2)a}$. We will call $\Delta\omega'^{\left(3\right)}=\chi'^{\left(3\right)}$ to compare it with the value that we have obtained: $\chi^{\left(3\right)}$.

This terms are not equal to the terms of equation (\ref{eq:deltaphi3}), but one must be careful with this expression. It was calculated using perturbation theory, taking $e$ as the eccentricity only for the first order therm, and as an initial condition of the equation of motion for higher terms. This is because $J$ in this works is defined as:

\begin{equation}
J^2=(1-e'^2)aGM
\end{equation}
for all terms. In equations (\ref{eq:angmom}), (\ref{eq:angmom1}) we can see what is the expression for $J^2$ in terms of $e$. So if we equate the two equations we can reach a relation between $e$ and $e'$:

\begin{equation}
(1-e'^2)=(1-e^2)\sum_{n=0}^\infty \left(\frac{r_{s}}{2a}\frac{3+e^{2}}{1-e^{2}}\right)^n
\end{equation}

So we can see that for $n=0$, $e=e'$. Now one can think that replacing the value of $e'$ in (\ref{eq:DoNhat}) will recover the first three terms of equation (\ref{eq:deltaphi3}), but that is not the case. The perturbation method consists in doing a iterative work, using the solution of the previous order to find the next order solution. The problem is that both works, in all iterations, neglect terms that will be important for the next iteration. For example, in the second order calculations, terms of third order in $\epsilon'$ are neglected, but this terms will contribute to the third order expression. If we replace the value of $e'$ in equation (\ref{eq:DoNhat}), we will recover the expansion:

\begin{equation}
\Delta\omega'^{\left(3\right)}= \chi'^{\left(3\right)}=3\pi\epsilon+\frac{3\pi\left(18+e^2\right)}{8}\epsilon^2+\frac{15\pi\left(15-6e-e^2-2e^3-e^4\right)}{16}\epsilon^3 \label{pert}
\end{equation}

As it can be seen, the last term of this expansion is not in agreement with that of equation (\ref{eq:deltaphi3}) in the third order. Furthermore, this term is smaller as $45(6/16)>15(15/16)$.

\section{Conclusions}
\label{Conclusions}

In this paper we have obtained higher order corrections (up to n-th order) for the perihelium precession using the Schwarzschild metric and complex integration, and
to compare it with different expansions appearing in the literature \cite{Fokas,Tyler,Rosales,Biesel,DEliseo,Scharf,Do_Nhat}, it was calculated the perihelion precession for the interior planets of the solar system and two hypothetical exoplanets around a star with the same mass of the Sun ($M_{\Theta}$). The values are shown in table (\ref{tab:planets}). Here we have used the notation $\Delta\omega' \left(\epsilon^{n}\right)$ or $\chi\left(\epsilon^{n}\right)$ for the contribution of the $n^{th}$ term and $\Delta\omega'^{\left(n\right)}$ or $\chi^{(n)}$ for the complete expansion until $n^{th}$ term.

\begin{table}[h]
\caption{Perihelion precession for the interior planets of the solar system and two hypotetical exoplanets. \label{tab:planets}}

\smallskip{}

\begin{tabular}{llllll}
\hline
System & Mercury & Venus & Earth & Exoplanet $\alpha$ & Exoplanet $\beta$\tabularnewline
\hline 
\hline  
$M(\times M_{\Theta})$ & 1.000 & 1.000 & 1.000 & 1.000 & 1.000 \tabularnewline
$r_s(UA)$ & $1.972\times 10^{-8}$ & $1.972\times 10^{-8}$ & $1.972\times 10^{-8}$ & $1.972\times 10^{-8}$ & $1.972\times 10^{-8}$ \tabularnewline
$a(UA)$ & 0.387 & 0.723 & 1.000 & 0.387 & 0.006\tabularnewline
$e$ & 0.206 & 0.007 & 0.017 & 0.950 & 0.200 \tabularnewline
$\epsilon$ & $5.319\times10^{-8}$ & $2.726\times10^{-8}$ & $1.972\times10^{-8}$ & $5.226\times10^{-7}$ & $3.423\times10^{-6}$\tabularnewline
$P(yr)$ & 0.241 & 0.615 & 1.000 & 0.240 & 0.006 \tabularnewline
\hline
$\chi(\epsilon)$ & $42.934''$ & $8.617''$ & $3.834''$ & $423.294''$ & $121738''$ \tabularnewline
$\chi(\epsilon^2)$ & $(5.149\times 10^{-6})''$ & $(5.286\times 10^{-7})''$ & $(1.702\times 10^{-7})''$ & $(5.227\times 10^{-4})''$ & $0.106''$ \tabularnewline
$\chi(\epsilon^3)$ & $(6.879\times 10^{-13})''$ & $(3.603\times 10^{-14})''$ & $(8.391\times 10^{-15})''$ & $(7.481\times 10^{-10})''$ & $(8.079\times 10^{-6})''$ \tabularnewline
$\chi^{(3)}$ & $42.934''$ & $8.617''$ & $3.834''$ & $423.294''$ & $121738,106''$ \tabularnewline
\hline
$\Delta\omega'(\epsilon)$ & $42.934''$ & $8.617''$ & $3.834''$ & $423.294''$ & $121738''$ \tabularnewline
$\Delta\omega'(\epsilon^2)$ & $(5.149\times 10^{-6})''$ & $(5.286\times 10^{-7})''$ & $(1.702\times 10^{-7})''$ & $(5.227\times 10^{-4})''$ & $0.106''$ \tabularnewline
$\Delta\omega'(\epsilon^3)$ & $(5.202\times 10^{-13})''$ & $(7.646\times 10^{-14})''$ & $(4.884\times 10^{-14})''$ & $(2.113\times 10^{-10})''$ & $(6.126\times 10^{-6})''$ \tabularnewline
$\Delta\omega'^{(3)}$ & $42.934''$ & $8.617''$ & $3.834''$ & $423.294''$ & $121738,106''$ \tabularnewline
\hline
\end{tabular}
\end{table}

It can be seen that the correction is not relevant. It is clear then that for weak gravitational fields, the classical formula is sufficiently precise. 

Now, to see if the correction is relevant for more massive objects, we calculated the perihelion precession for three binary systems, Sagittarius A*-S2, OJ287 and H1821+643. In table (\ref{tab:Oj287}) are shown this calculations performed using equations  (\ref{eq:ang2}),(\ref{eq:chi3}) and (\ref{pert})  for different orders.

Sagitarius A* is a bright and very compact radio source located at the center of the Milky Way. It is theorized that Sagittarius A* is a supermassive black hole. We took star S2, because is the one that presents a very peculiar orbit. As it can be seen in table (\ref{tab:Oj287}), the corrections are more significant that those for the Solar System planets.

\begin{table}[h]
\caption{Perihelion advance in degrees per period for some binary systems. \label{tab:Oj287}}
\smallskip{}

\centering{}%
\begin{tabular}{llllll}
\hline
System & Sagittarius A*-S2  & OJ287 & H1821+643 \tabularnewline
\hline
\hline  
$M(\times M_{\Theta})$ & $\left(4.310\right)\times10^{6}$ & $1.830\times10^{10}$ & $3.000\times10^{10}$ \tabularnewline
$r_s(AU)$ & 0.085 & 360.847 & $591.553$ \tabularnewline
$a(AU)$ & $923.077$ & $11500$ & 40000 \tabularnewline
$e$ & $0.870$ & $0.700$ & 0.900 \tabularnewline
$\epsilon$ & $3.787\times 10^{-4}$ & $6.153\times 10^{-2}$ & $7.784\times 10^{-2}$ \tabularnewline
\hline
$\chi(\epsilon)$ & $0.205^\circ$ & $33.223^\circ$ & $42.031^\circ$ \tabularnewline
$\chi(\epsilon^2)$ & $(1.816\times 10^{-4})^\circ$ & $4.724^\circ$ & $7.692^\circ$ \tabularnewline
$\chi(\epsilon^3)$ & $(1.858\times 10^{-7})^\circ$ & $0.765^\circ$ & $1.625^\circ$ \tabularnewline
$\chi(\epsilon^4)$ & $(2.058\times 10^{-10})^\circ$ & $0.133^\circ$ & $0.373^\circ$ \tabularnewline
$\chi^{(3)}$ & $0.205^\circ$ & $38.713^\circ$ & $51.349^\circ$ \tabularnewline
$\chi^{(4)}$ & $0.205^\circ$ & $38.846^\circ$ & $51.722^\circ$ \tabularnewline
\hline
$\Delta\omega'(\epsilon)$ & $0.205^\circ$ & $33.223^\circ$ & $42.031^\circ$ \tabularnewline
$\Delta\omega'(\epsilon^2)$ & $(1.816\times 10^{-4})^\circ$ & $4.724^\circ$ & $7.692^\circ$ \tabularnewline
$\Delta\omega'(\epsilon^3)$ & $(6.539\times 10^{-8})^\circ$ & $0.369^\circ$ & $0.531^\circ$ \tabularnewline
$\Delta\omega'^{(3)}$ & $0.205^\circ$ & $38.317^\circ$ & $50.255^\circ$ \tabularnewline
\hline
\end{tabular}
\end{table}

OJ 287 is a binary system that produces periodic outbursts. This outbursts have been detected for approximately 100 years. This system is located 3.500 million light years from Earth, and is theorized that it is a binary system of black holes having a total mass of around $1.845\times 10^{10}M_{\Theta}$. It can be seen that between the first and the second order terms ($\chi\left(\epsilon\right)$ and $\chi\left(\epsilon^{2}\right)$ or $\Delta\omega'\left(\epsilon\right)$ and $\Delta\omega'\left(\epsilon^{2}\right)$)  there is a difference of approximately $5^{\circ}$. For higher orders, the difference is less than $1^{\circ}$. At the third order the contribution to  the perihelion precession is $0.765^{\circ}$ in our expansion and $0.369^{\circ}$ in the corresponding expansions calculated by D'Eliseo and Nhat. Something interesting about our expansion is that it begins to stabilize taking into account higher order corrections around $38.8^{\circ}$. This  is in good agreement with the  measured  value that is approximately $39^{\circ}$ in a cycle.Then, equation (\ref{eq:ang2}) gives an important correction to the perihelion precession.The calculation can be improved if it is considered the spin of the central black hole, and the gravitational radiation of the system. Also it is important to recall that the experimental measurements have been questioned because of the limited number of orbital companions of the system.

The last system is H1821+643, that corresponds to the most massive black hole ever detected. With a mass of $3\times 10^{10}M_{\Theta}$, the orbital parameters of the gravitational companion are not known, so we used random parameters. Also, for this two super-massive systems the corrections are relevant.

With all this results, it can be concluded that the approach presented in this paper is the best to calculate the perihelion precession for binary systems as it can be calculated until any order in the third root of the motion equation $R_o$ (see equation (\ref{eq:raiz})). With other methods as perturbation theory, this calculation is more difficult and there are some approximations that must be made to work with this method that can carry errors.


\section{Orbit profile}

Using Verlet's method for solving differential equations, one can simulate the shape of the relativistic orbits. It is not recommended to use Euler's method because it does not conserve the energy and the orbits will have a spiral shape.  

The parameters for the simulation are the central mass of the system $M$, the eccentricity $e$ and the semi-mayor axis $a$. With this we can  calculate the energy and the angular momentum. Then we proceed to divide the motion in $n$ steps in time as: $\tau_n=\tau_o+n\Delta\tau$, where $\Delta\tau$ is the time variation that must be small.

Now, from equation (\ref{eq:mov}) we can compute the first two derivatives of $r(\tau)$:

\begin{equation}
\frac{dr}{d\tau}=\sqrt{A+\frac{r_sc^2}{r}-\frac{J^{2}}{r^{2}}+\frac{J^{2}r_s}{r^3}}
\end{equation}

\begin{equation}
\frac{d^2r}{d\tau^2}=-\frac{r_sc^2}{2r^2}+\frac{J^{2}}{r^3}-\frac{3J^{2}r_s}{2r^4}
\end{equation}

Then, we use Euler's method only for the first step:

\begin{equation}
r_{1}=r_o+\frac{dr}{d\tau}\vert_{r_o}\Delta\tau+\frac{1}{2}\frac{d^2r}{d\tau^2}\vert_{r_o}\Delta\tau^2
\end{equation}
After this step, we employ  Verlet's integration method:

\begin{equation}
r_{n+1}=2r_{n}-r_{n-1}+\frac{d^2r}{d\tau^2}\vert_{r_n}\Delta\tau^2
\end{equation}

Also, in every step, it is necessary to use equation (\ref{eq:angmom}) to integrate $\phi$:

\begin{equation}
\phi_{n+1}=\phi_n+\frac{d\phi}{d\tau}\vert_{t_n}\Delta\tau=\phi_n+\frac{J}{r_n^2}\Delta\tau
\end{equation}

\begin{figure}[htpb]
\includegraphics[scale=0.5]{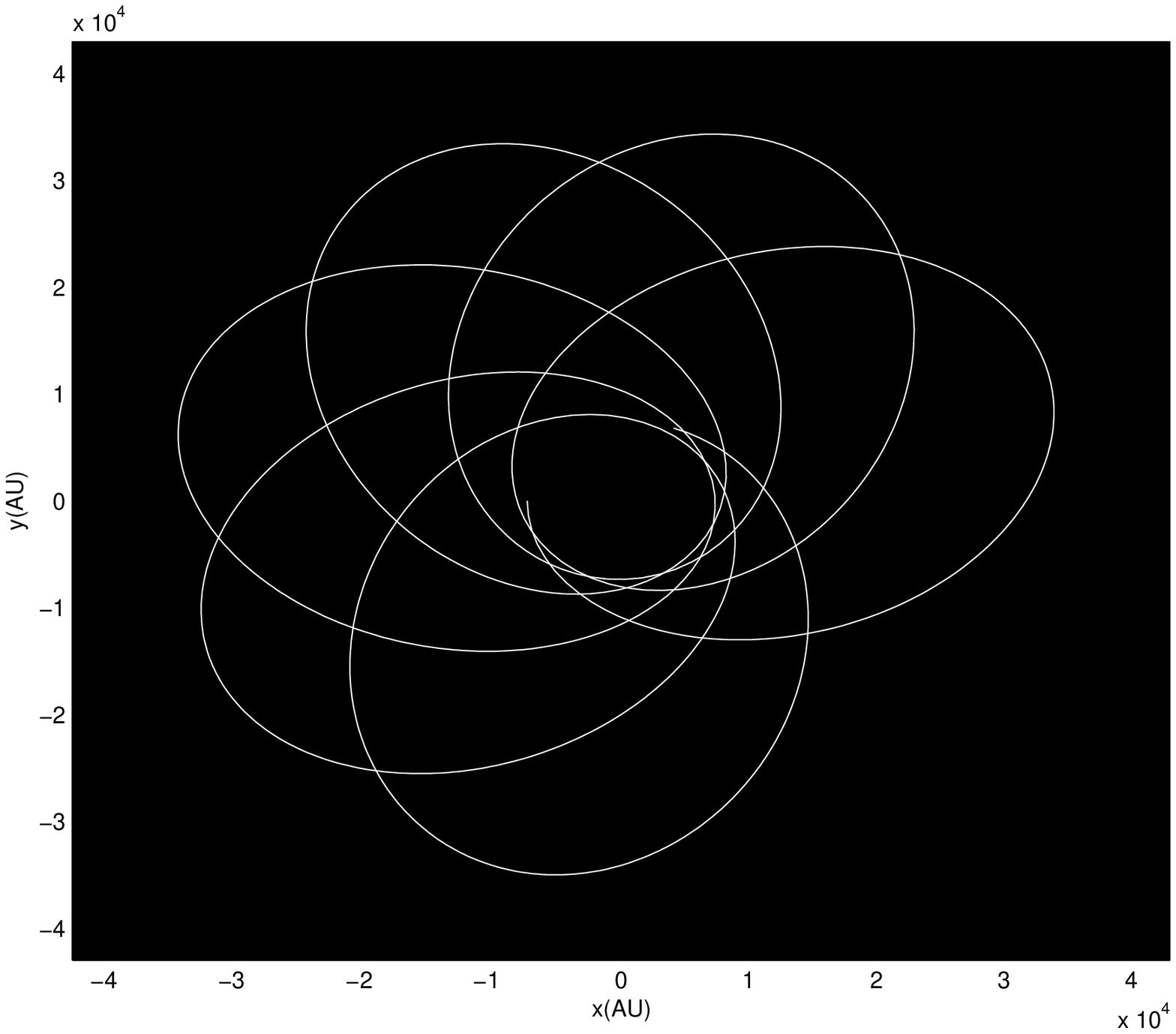}
\includegraphics[scale=0.5]{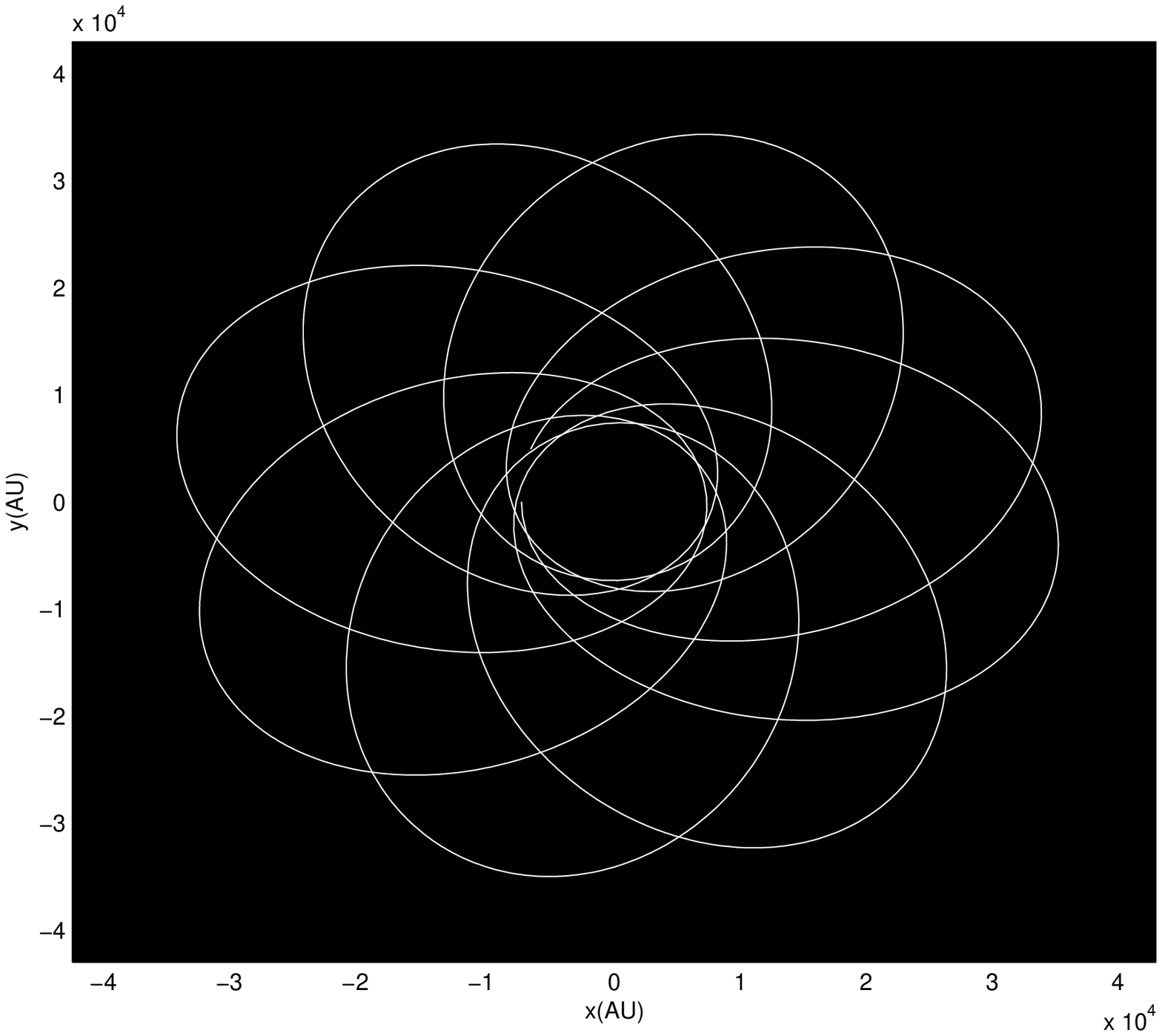}
\caption{Trajectory of a system with central mass $M = 4\times 10^{10} M_{\Theta}$, eccentricity $e=0.658$ and $a=21500 AU$.\label{fig:avance-1-2}}
\end{figure}

\begin{figure}[htpb]
\includegraphics[scale=0.5]{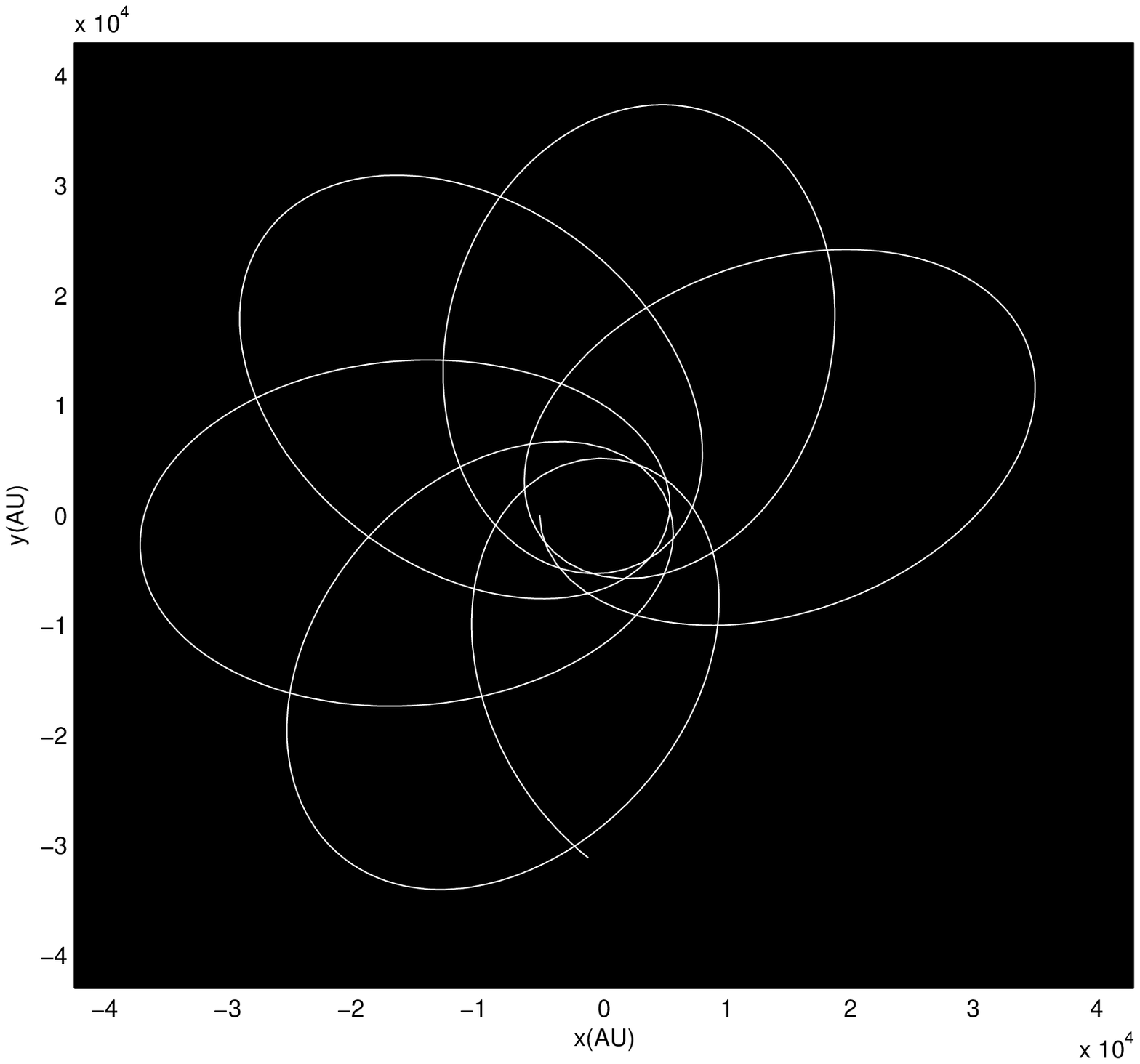}
\includegraphics[scale=0.5]{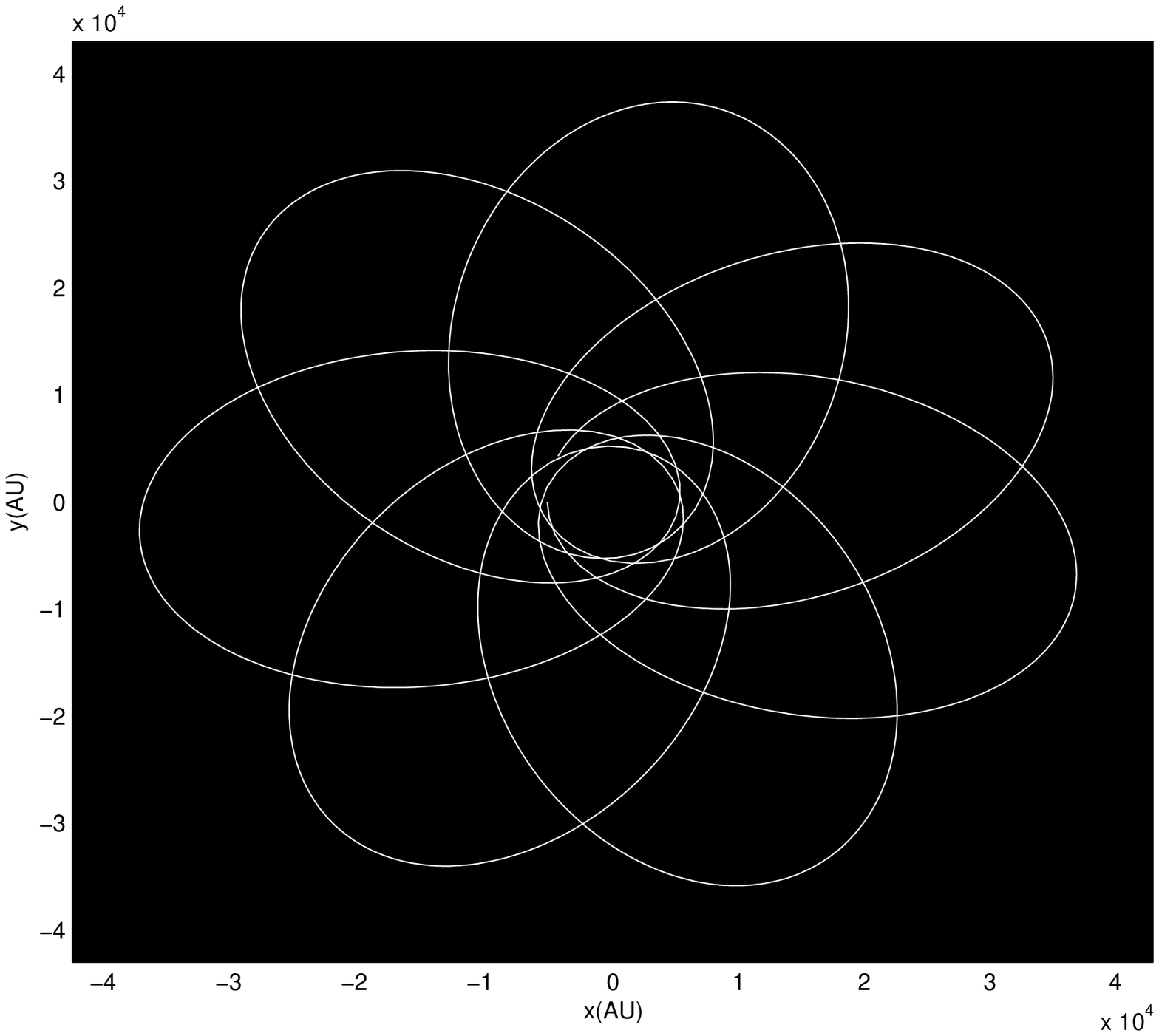}
\caption{Trajectory of a system with central mass $M = 3.3\times 10^{10} M_{\Theta}$, eccentricity $e=0.58$ and $a=21500 AU$.\label{fig:avance-1-3}}
\end{figure}

\begin{figure}[htpb]
\includegraphics[scale=0.5]{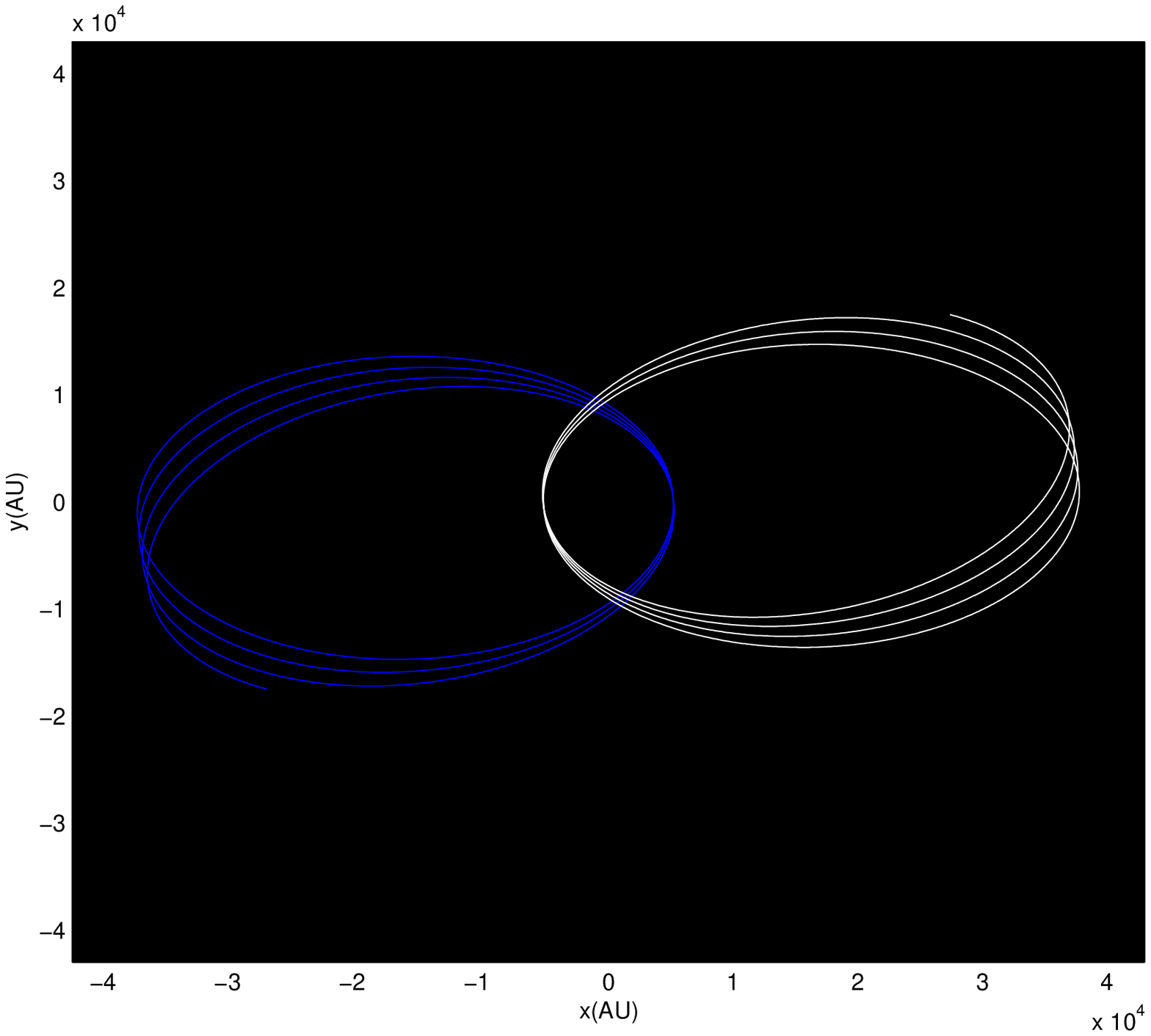}
\includegraphics[scale=0.5]{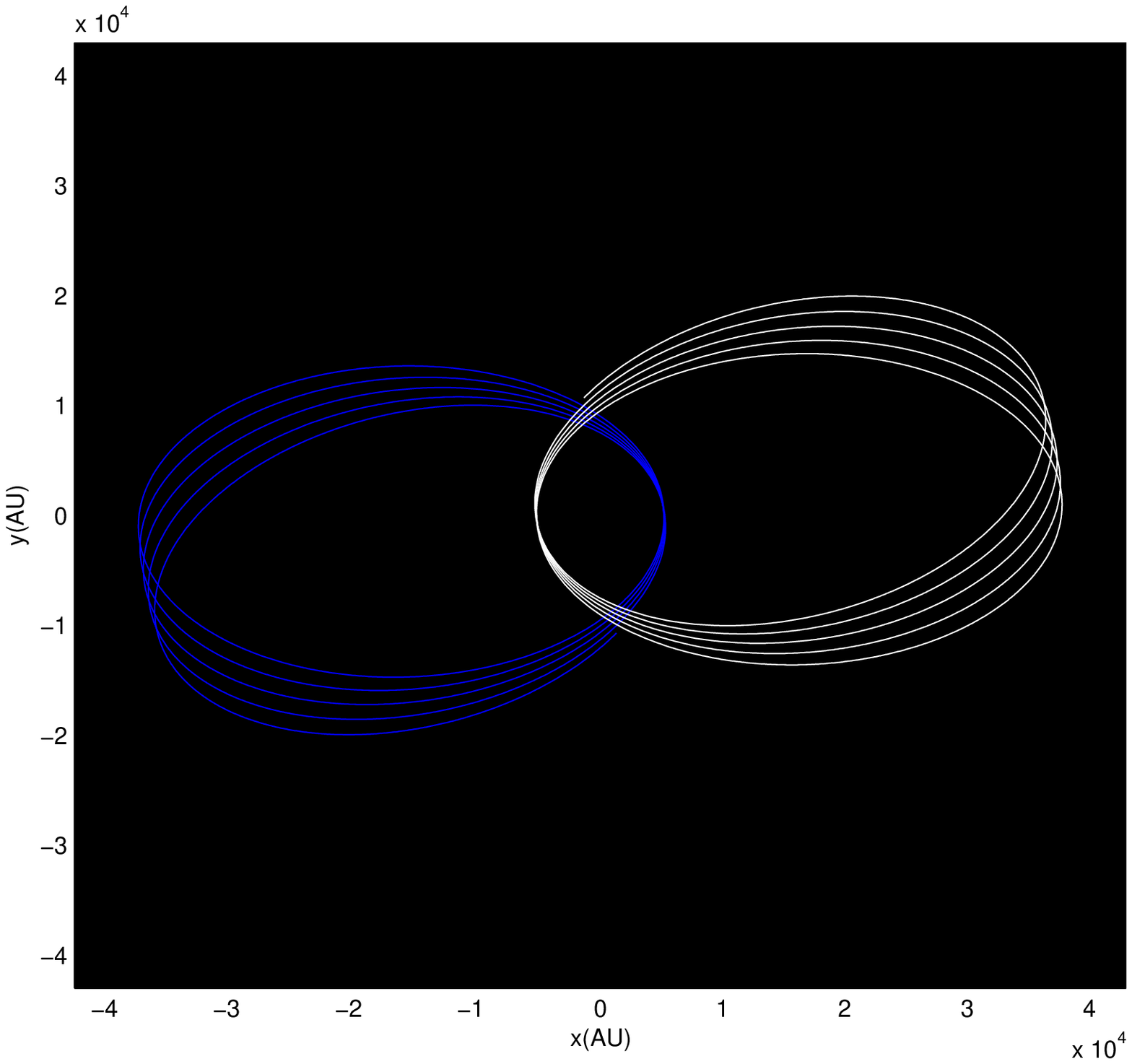}
\caption{Trajectory of a binary system with equal masses of $1.65\times 10^9 M_{\Theta}$. The parameters are $e=0.758$, $a=21500 AU$.\label{fig:avance-1-4}}
\end{figure}

Finally, with the solutions we have generated figures  \ref{fig:avance-1-2}, \ref{fig:avance-1-3} and \ref{fig:avance-1-4} for different binary systems.

\newpage

\begin{figure}[htpb]
\includegraphics[scale=0.5]{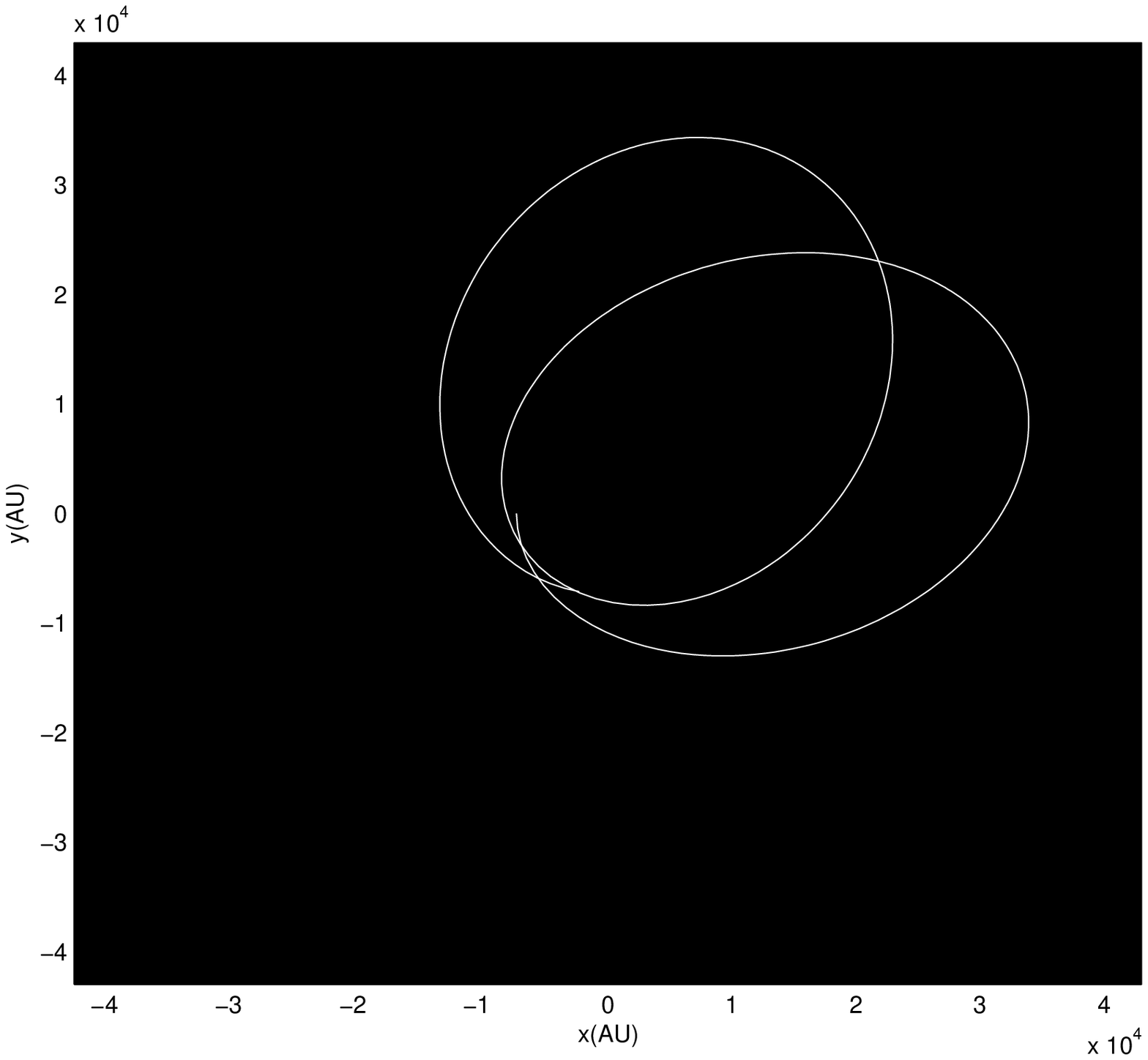}
\includegraphics[scale=0.5]{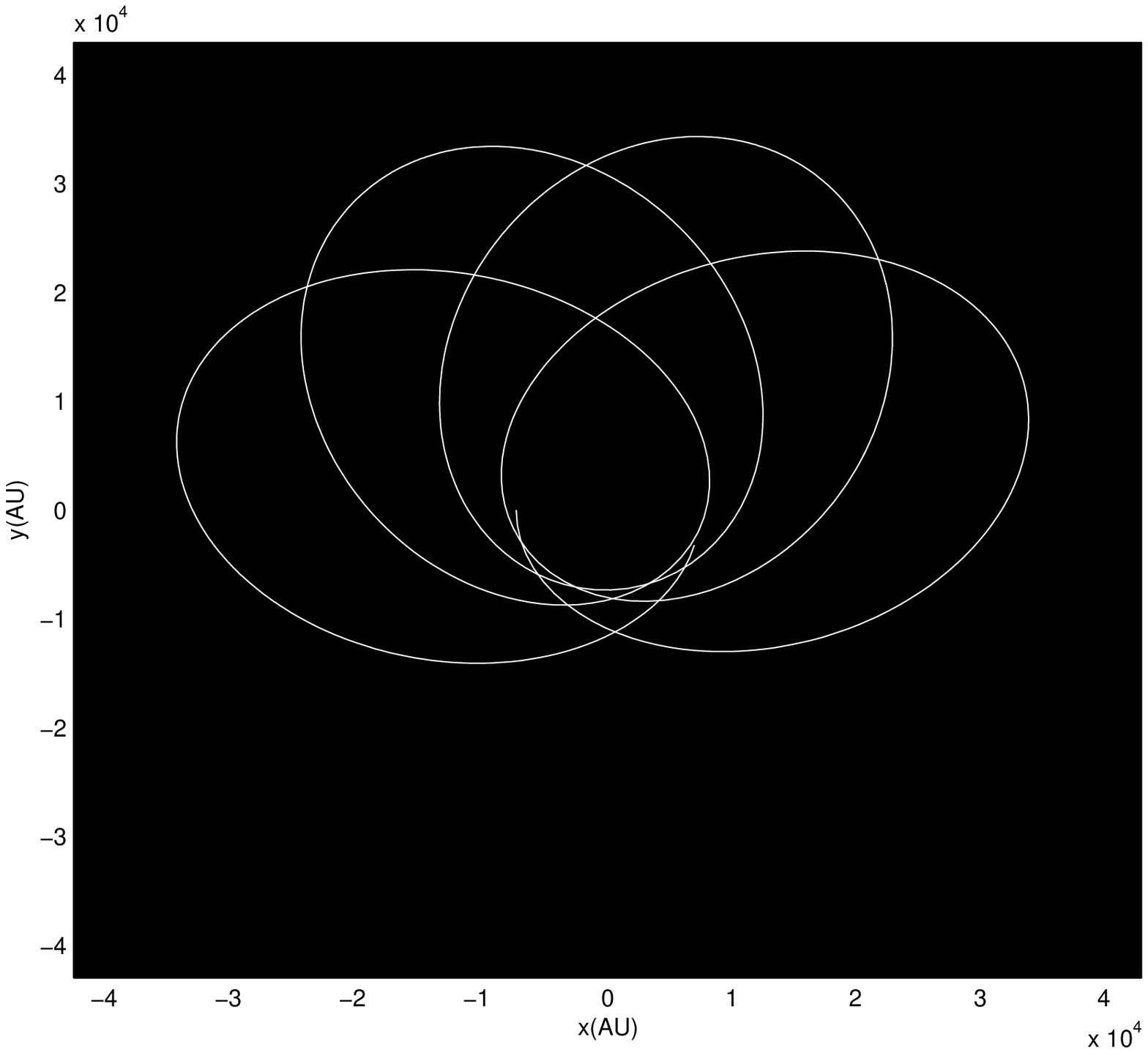}
\end{figure}
\newpage
\newpage
\newpage

\begin{figure}[htpb]
\includegraphics[scale=0.5]{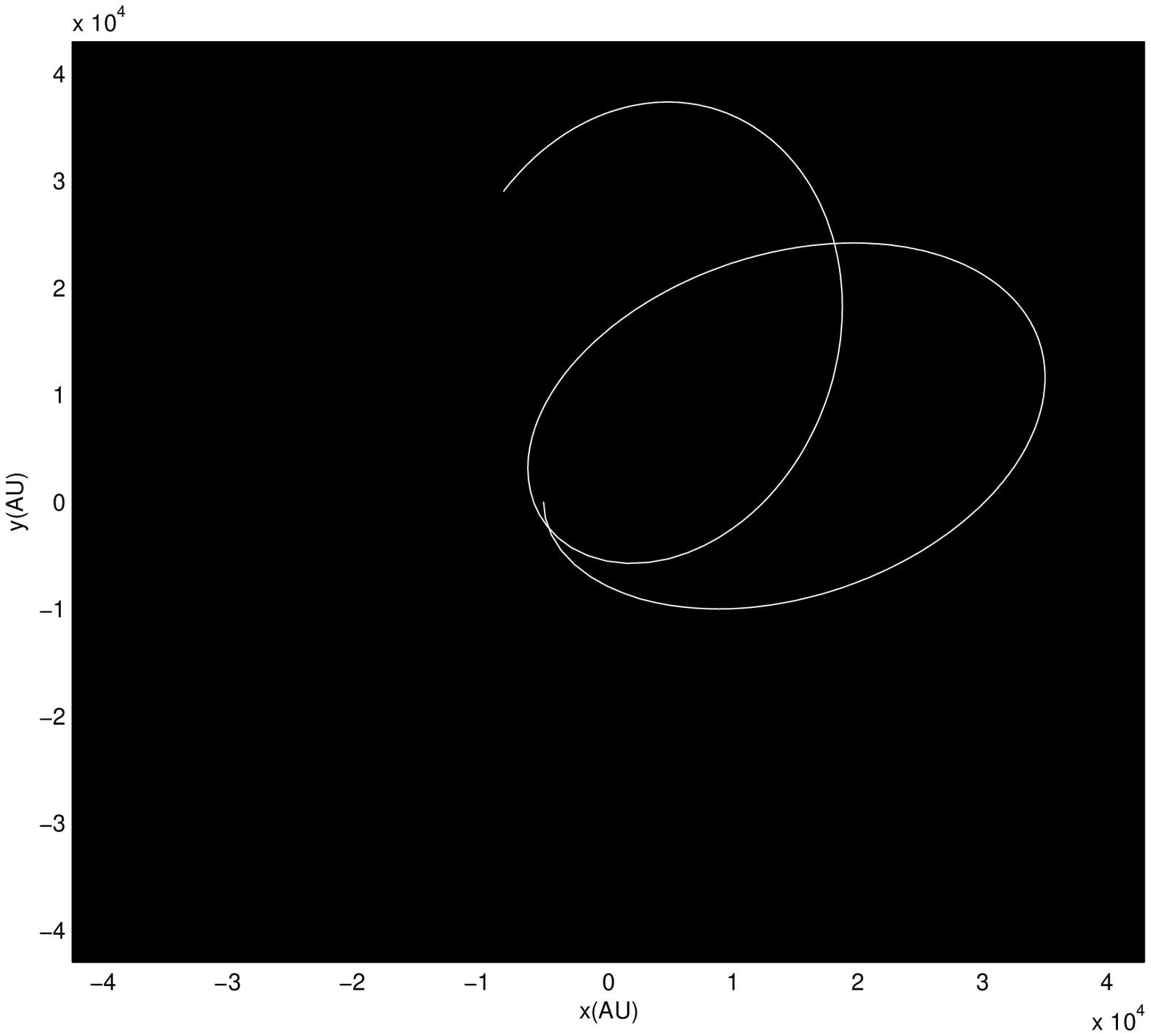}
\includegraphics[scale=0.5]{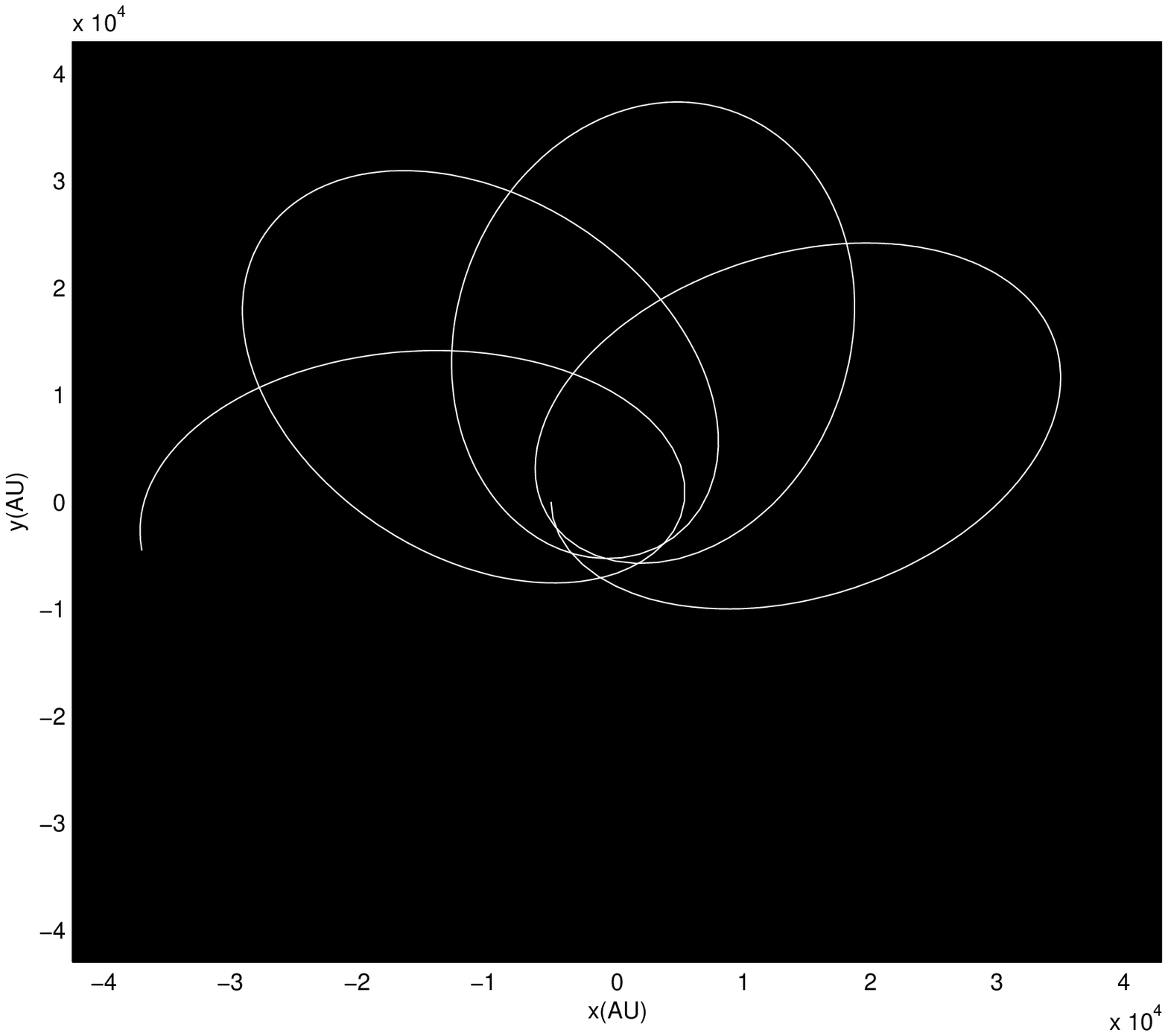}
\end{figure}
\newpage
\newpage
\newpage

\begin{figure}[htpb]
\includegraphics[scale=0.5]{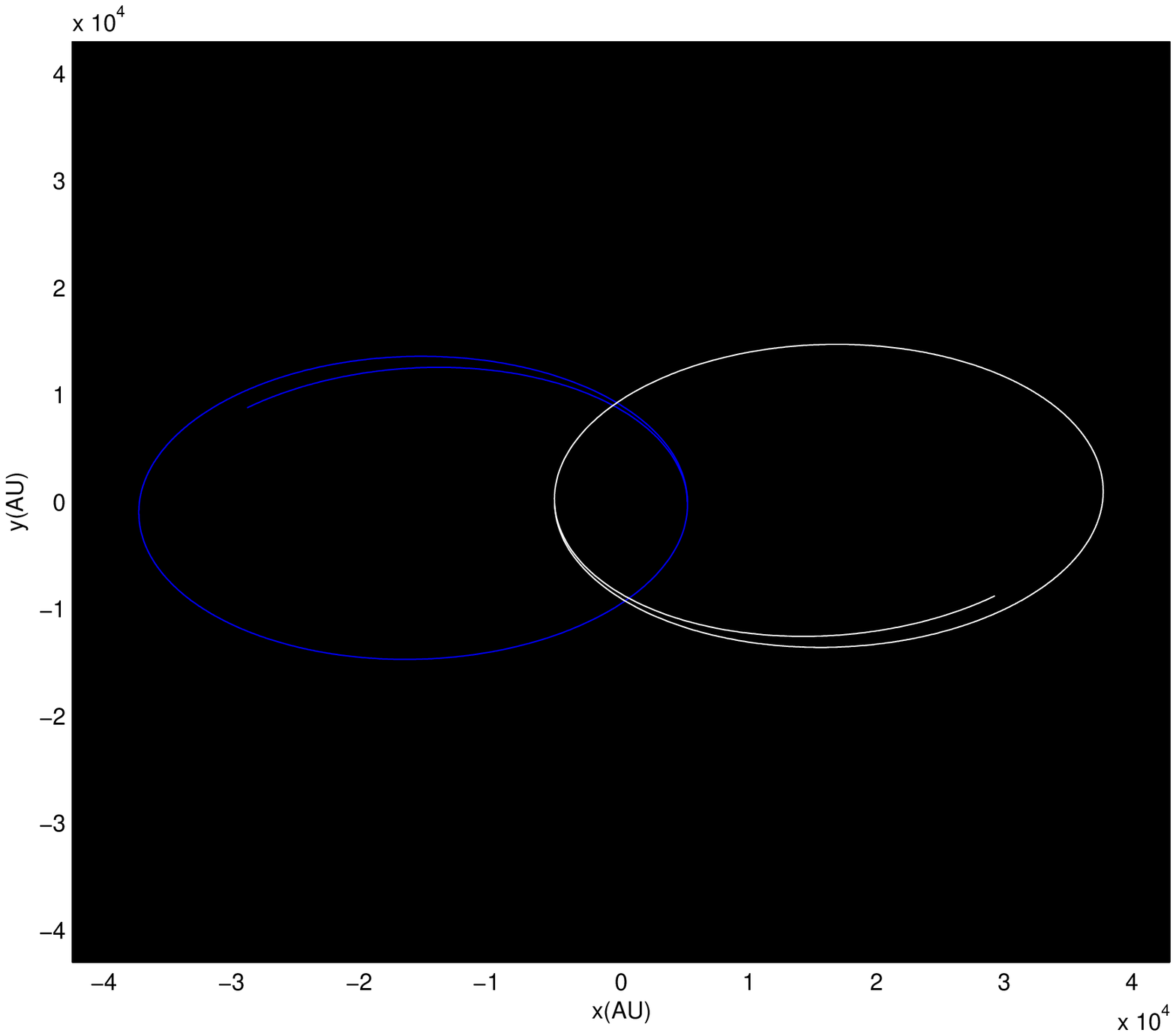}
\includegraphics[scale=0.5]{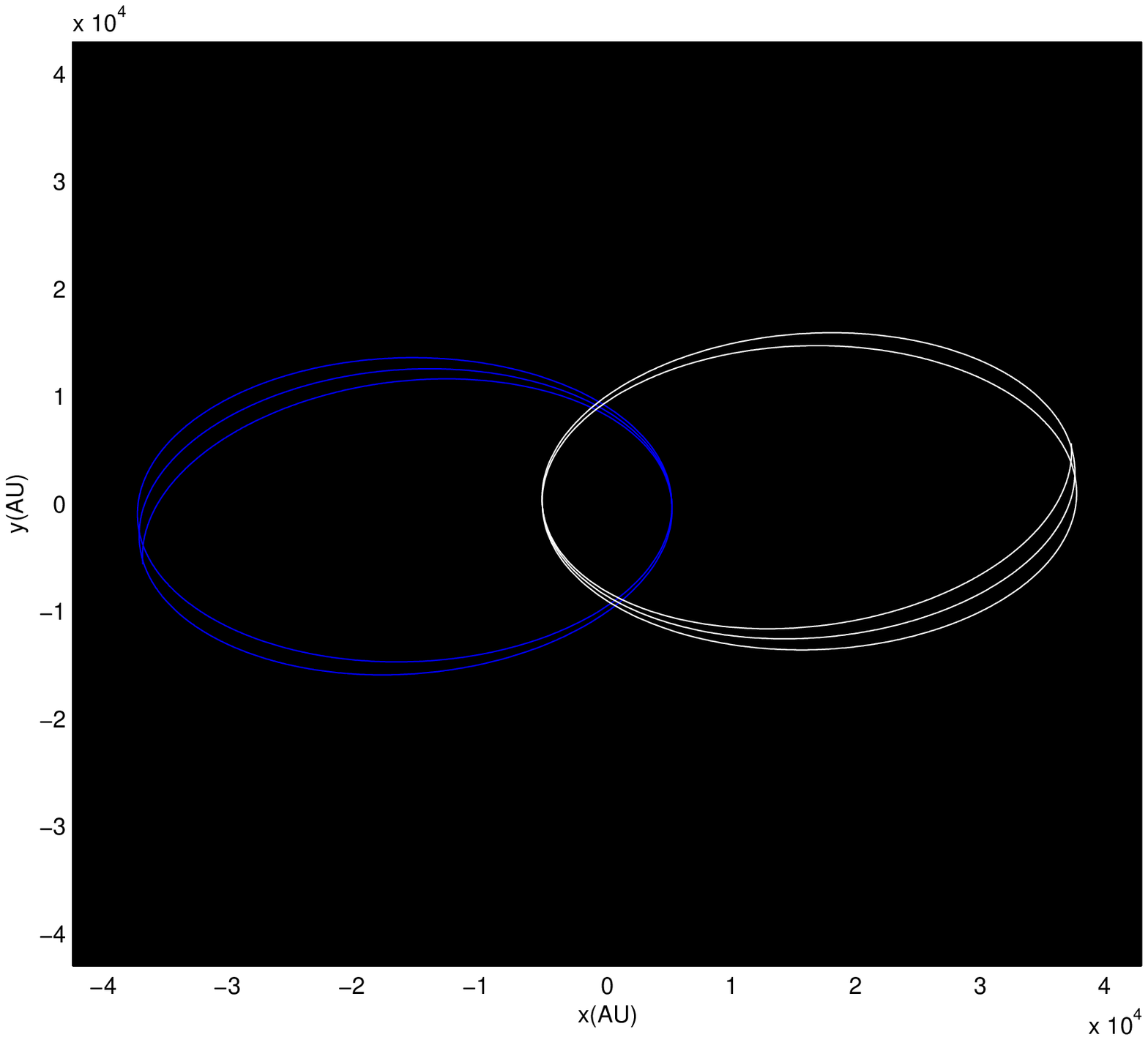}
\end{figure}

\end{document}